\documentclass[]{aa}
\usepackage{graphicx}
\usepackage{subfig}
\usepackage{multirow}
\usepackage{hyperref}
\hypersetup
{
    colorlinks= true,
    linkcolor = blue,
    filecolor = blue,      
    urlcolor  = blue,
    citecolor = blue
}
\usepackage{etoolbox}
\makeatletter
\patchcmd{\NAT@citex}
  {\@citea\NAT@hyper@{%
     \NAT@nmfmt{\NAT@nm}%
     \hyper@natlinkbreak{\NAT@aysep\NAT@spacechar}{\@citeb\@extra@b@citeb}%
     \NAT@date}}
  {\@citea\NAT@nmfmt{\NAT@nm}%
   \NAT@aysep\NAT@spacechar\NAT@hyper@{\NAT@date}}{}{}

\patchcmd{\NAT@citex}
  {\@citea\NAT@hyper@{%
     \NAT@nmfmt{\NAT@nm}%
     \hyper@natlinkbreak{\NAT@spacechar\NAT@@open\if*#1*\else#1\NAT@spacechar\fi}%
       {\@citeb\@extra@b@citeb}%
     \NAT@date}}
  {\@citea\NAT@nmfmt{\NAT@nm}%
   \NAT@spacechar\NAT@@open\if*#1*\else#1\NAT@spacechar\fi\NAT@hyper@{\NAT@date}}
  {}{}
\makeatother
\usepackage{txfonts}
\usepackage{natbib}
\bibpunct{(}{)}{;}{a}{}{,} 
\def\app#1#2{%
  \mathrel{%
    \setbox0=\hbox{$#1\sim$}%
    \setbox2=\hbox{%
      \rlap{\hbox{$#1\propto$}}%
      \lower1.1\ht0\box0%
    }%
    \raise0.25\ht2\box2%
  }%
}

\begin{document}

   \title{Water emission tracing active star formation from the Milky Way to high-\textit{z} galaxies}

   \author{K. M. Dutkowska
          \inst{}
          \and
          L. E. Kristensen
          \inst{}
          }

   \institute{Niels Bohr Institute \& Centre for Star and Planet Formation, Copenhagen University,
              \O{}ster Voldgade 5-7, 1350 Copenhagen K, Denmark\\
              \email{dutkowska@nbi.ku.dk}
             }

   \date{Received xxx xxx, xxxx; accepted xxx xx, xxxx}

 
  \abstract
    {The question of how most stars in the Universe form remains open. While star formation predominantly takes place in young massive clusters, the current framework focuses on isolated star formation. This poses a problem when trying to constrain the initial stellar mass and the core mass functions, both in the local and distant Universe.}
    {One way to access the bulk of protostellar activity within star-forming clusters is to trace signposts of active star formation with emission from molecular outflows. These outflows are bright, e.g., in water emission, which is observable throughout cosmological times, providing a direct observational link between nearby and distant galaxies. We propose to utilize the in-depth knowledge of local star formation as seen with molecular tracers, such as water, to explore the nature of star formation in the Universe.}
    {We present a large-scale statistical galactic model of emission from galactic active star-forming regions. Our model is built on observations of well-resolved nearby clusters. By simulating emission from molecular outflows, which is known to scale with mass, we create a proxy that can be used to predict the emission from clustered star formation at galactic scales. In particular, the para-H$_2$O \(2_{02} - 1_{11}\) line is well-suited for this purpose, as it is among one of the brightest transitions observed toward Galactic star-forming regions and is now routinely observed toward distant galaxies.}
    {We evaluated the impact of the most important global-star formation parameters (i.e., initial stellar mass function, molecular cloud mass distribution, star formation efficiency, and free-fall time efficiency) on simulation results. We observe that for emission from the para-H$_2$O \(2_{02} - 1_{11}\) line, the initial mass function and molecular cloud mass distribution have a negligible impact on the emission, both locally and globally, whereas the opposite holds for star-formation efficiency and free-fall time efficiency. Moreover, this water transition proves to be a low-contrast tracer of star formation, with $\int I_{\nu}\propto {M_\mathrm{env}}$.}
    {The fine-tuning of the model and adaptation to morphologies of distant galaxies should result in realistic predictions of observed molecular emission and make the galaxy-in-a-box model a tool to analyze and better understand star formation throughout cosmological times.}
    

   \maketitle
%

\section{Introduction}
  
  Water is one of the key molecules tracing active and current star formation (SF); in the Milky Way water emission is almost uniformly associated with molecular outflows from protostars \citep{2021A&A...648A..24V}. These outflows arise at the earliest stages of star formation, when the protostar is in its main accretion phase and the interaction between the infalling envelope, winds and jets launched from the protostar is particularly strong \citep{2016ARA&A..54..491B}. When this happens, water, predominantly locked up as ice on dust grains, is released from the icy grain mantles into the gas phase, causing a jump in the abundance of many orders of magnitude. At the same time, the physical conditions are conducive to water being readily excited into rotational states, and the de-excitation leads to subsequent cooling \citep{2014MNRAS.440.1844S}. Therefore, whenever star formation occurs, these outflows light up in water emission.
  
  Water emission is also observed towards high-redshift galaxies \citep[e.g.,][]{2016A&A...595A..80Y,2019ApJ...880...92J,2021A&A...646A.178S}. The origin of this emission is interpreted to be the molecular clouds from which stars form, and not the protostellar outflows. This interpretation is primarily grounded in a very tight correlation between the far-infrared luminosity ($L_{\mathrm{FIR}}$) and water line luminosity ($L_{\mathrm{H}_2\mathrm{O}}$), where $L_{\mathrm{FIR}}$ is thought to trace dust \citep[e.g.,][]{2008ApJ...675..303G,2014A&A...567A..91G,2013A&A...551A.115O}. The latter indicates that $L_{\mathrm{FIR}}$ indirectly traces molecular clouds, and the excitation of water molecules is expected to be caused by the FIR radiation field through radiative pumping.
  
  Two dominant mechanisms contribute to returning the water ice into the gas phase. The first, and the most effective, is thermal desorption if the temperature of the dust grains rises above $\sim100$~K \citep[e.g.,][]{2001MNRAS.327.1165F}. Such high temperatures are typically found within the inner $\sim$ 10$^2$ AU of forming stars \citep[e.g.,][]{2007A&A...465..913B}. The second is sputtering of ice from the dust grains when neutral species or ions with sufficient kinetic energy (predominantly H$_2$, H and He) collide with the ice mantle. Due to its highly energetic character, sputtering can cause the dissociation of water molecules. However, the high temperatures within outflows make the gas-phase synthesis of water effective enough to sustain the high abundance of water molecules \citep{2014MNRAS.440.1844S}. Finally, water may also be directly synthesized in the gas from ion-neutral reactions. In dark molecular clouds, this path is inefficient \citep{2009ApJ...690.1497H}, but in photon and X-ray-dominated regions (PDRs and XDRs) where the ionization fraction is high, this mechanism may be the dominant \citep{2011A&A...525A.119M}. 
  
  Observations of emission from the ground state levels of ortho- and para-water, e.g., the ortho-H$_2$O $1_{10} - 1_{01}$ line at 557 GHz, are known to trace the warm outflowing gas \citep{2014A&A...572A..21M}, as do the mid-excited transitions, with $E_\mathrm{up} \sim100 - 300$ K, like the para$-$H$_2$O $2_{02} - 1_{11}$ line at 988 GHz. Subsequently, highly excited water transitions with $E_\mathrm{up} > 300$ K, such as the ortho-H$_2$O $5_{23} - 5_{14}$ line at 1411 GHz, are only populated in high-temperature gas and strong shocks \citep{2013ChRv..113.9043V}. Water, except for the ground state transitions, may also be excited by pumping to higher-excited levels by FIR photons \citep{2014A&A...567A..91G}. However, in the Galactic outflows where water excitation is collisionally dominated, there are no signs that other processes, such as FIR pumping, play any significant role in the excitation \citep{2014A&A...572A..21M}. It poses a question: does water behave differently at high redshift?
  
  With the great progress in astrochemistry in the past years, particularly thanks to the observational programs carried out with the \textit{Herschel} Space Observatory (active between $2009-2013$) and the Atacama Large Millimeter/submillimeter Array (ALMA), we are now routinely observing the distant Universe in molecular line emission \citep{2020RSOS....700556H}. Numerous surveys provided detailed chemical inventories of star-forming regions within the Galaxy \citep[for a recent review, see][]{2020ARA&A..58..727J}, and as we observe the same molecules across the Universe \citep{2021arXiv210913848M}, we can now start to fill the informational gap between high-redshift galaxies and the Milky Way and start comparing the observational results between these regimes.
  
  One of the questions we can answer is, how molecular line emission can be used to quantitatively trace active star formation? Most stars form in clusters \citep{2003ARA&A..41...57L}. In clusters all ranges of stellar masses are present and relatively few main-sequence high-mass stars can easily outshine the entire low-mass population. Moreover, the younger the protostar, the deeper it is embedded in gas and dust. Therefore, we need to use reliable tracers of active star formation that are common and bright enough to be easily observed. One of the best tracers in our Galaxy, also observed in the distant Universe, is water: the emission is particularly bright in the deeply embedded phase, when the protostars drive molecular outflows \citep[e.g.,][]{2016ARA&A..54..491B}. 
  
  In this work, we present a model, which can be used to compare observations from different galaxies with the emission that could arise from active star-forming regions. In the model, we simulate emission from molecular outflows, one of the key signposts of active and current star formation, that would arise from protostars within star-forming clusters. These star-forming clusters are then incorporated into a large-scale galactic model, which contains a range of molecular clouds in which the stars form. In this study we focus on simulating water emission at 988 GHz (the $J_{KaKc} = 2_{02} - 1_{11}$ line), which is particularly bright in Galactic star-forming regions and has been observed towards many high-redshift galaxies \citep[e.g.,][]{2021A&A...648A..24V,2013A&A...554A..83V}, but the model is set up such that it can ingest and predict any type of outflow emission.
  
  This paper is organized as follows. Section \ref{model} describes our galactic model in detail and provides the methods used to obtain the results. Subsequently, in Section \ref{results} we present the results of a parameter space study of the model, which we then discuss and present future prospects for in Section \ref{discussion}. Finally, we present our conclusions in Section \ref{conclusions}.
   
\section{Model}
\label{model}
    
\begin{figure}[t!]
      \resizebox{\hsize}{!}{\includegraphics{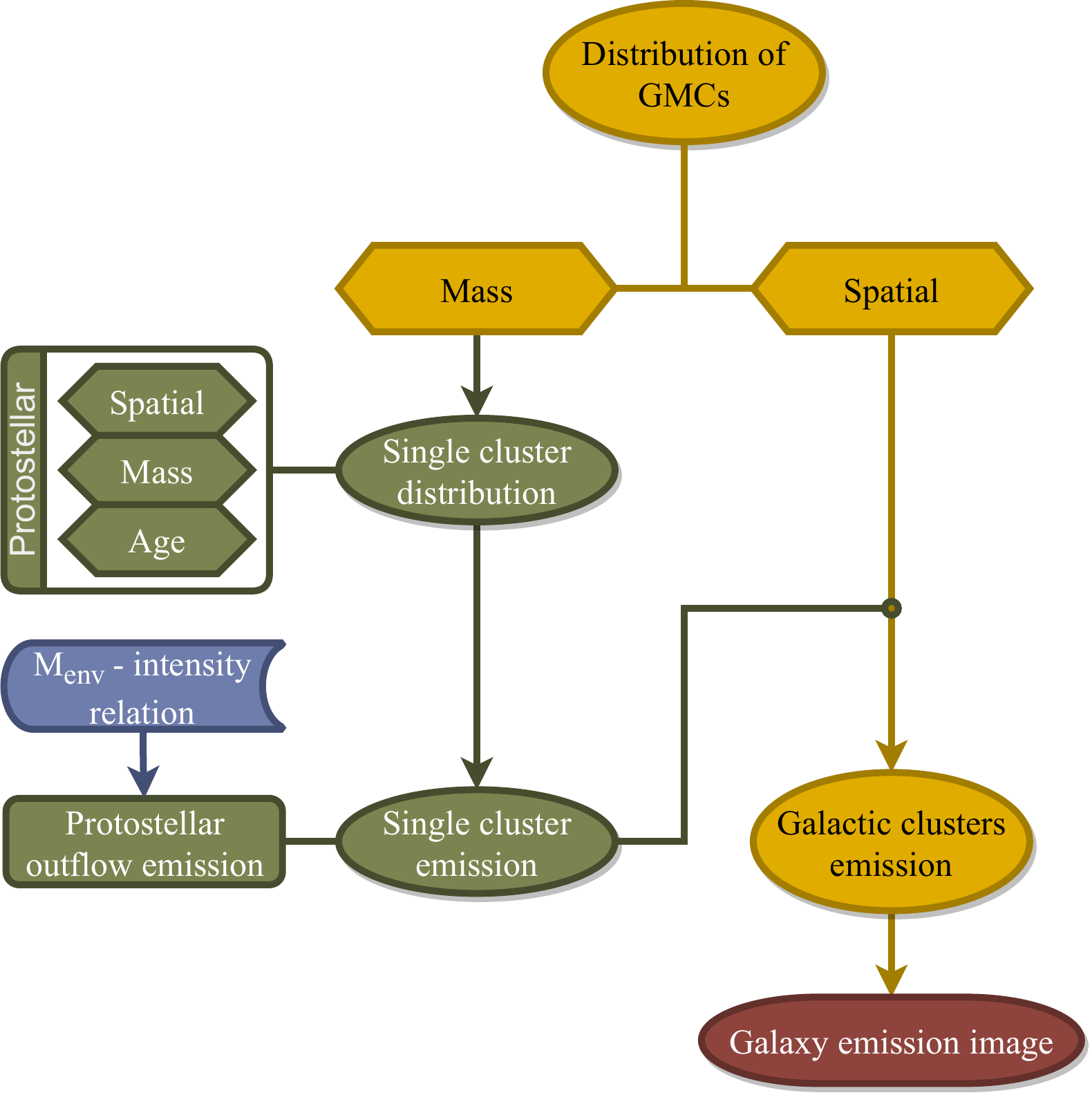}}
      \caption{Schematic flowchart of the galaxy-in-a-box model. Computations starts with generating the spatial  (Sect. \ref{spat}) and mass (Sect. \ref{mcmmass}) distributions of GMCs in the simulated galaxy. The GMC mass distribution serves as the input to the module generating the protostellar spatial, mass, and age distributions within individual star-forming clusters. Here, each randomly chosen GMC mass is an initial mass of the cluster. Having calculated these distributions, the model uses them to assign molecular outflow emission to each protostar within the cluster, based on the envelope mass - outflow intensity relation calculated using the Water Emission Database (Sect. \ref{wed}). After repeating these calculations for all GMCs, the emission and mass of star-forming clusters are returned to the galaxy-in-a-box model. Subsequently, the model merges the spatial distribution of the initial GMCs with water emission emerging from the corresponding star-forming clusters. Once the model returns the expected emission from the galaxy, this raw galactic emission grid is convolved with a Gaussian beam, producing the integrated intensity image of the galaxy. In this flowchart, yellow corresponds to the galactic part of the model (Sect. \ref{giab}), green to the cluster model (Sect. \ref{ciab}), red is for the last stage of the model, and blue indicates the external data input.}
    \label{flowchart}
    \end{figure}

    On galactic scales, stars predominantly form in Giant Molecular Clouds (GMCs). These GMCs form complexes, which follow a certain spatial distribution in galaxies, as will be outlined below. Hence, to build a model of galactic emission from active star-forming regions, we broke this distribution down into its constituent parts. We used an existing cluster model (Sect. \ref{ciab}) as a starting point and adapted it into a cloud model. We subsequently used this cloud model as the building blocks for the galaxy-in-a-box model (see Sect. \ref{giab}). Finally, we built the observational template used for emission assignment in the form of a database in which we gathered the available water data from ground-based observations and the \textit{Herschel} Space Observatory (Sect. \ref{wed}). The model is outlined in Fig. \ref{flowchart} with the different modules highlighted.
    
\subsection{Cluster-in-a-box model}
\label{ciab}

    Most stars form in clusters, especially in high-mass clusters \citep{2003ARA&A..41...57L}. These clusters harbor protostars covering the whole range of stellar masses. However, at the time of formation they are also deeply embedded in their natal clouds, and so it is impossible to access the initial main-sequence stellar populations forming within these clusters directly. Moreover, massive stars dominate cluster emission, making the low-mass population hard to access observationally. An alternative is to probe this population with outflow emission. Studies show that there is a proportionality between this emission and protostellar envelope mass \citep[e.g.,][]{1996A&A...311..858B,2022skretas}. \cite{2015ApJ...807L..25K} utilized this link to construct the cluster-in-a-box model\footnote{The codes required for running the model are publicly available: \url{https://github.com/egstrom/cluster-in-a-box} \\ \href{https://doi.org/10.5281/zenodo.13184}{doi: 10.5281/zenodo.13184}}, simulating methanol emission from low-mass outflows in embedded star-forming clusters. 
 
    The cluster model consists of a template cluster and molecular emission assigned to each protostar in the cluster. The spatial distribution of protostars in the template cluster is based on the model by \cite{2014ApJ...789...86A}, where the radial extent of the cluster can be described by the power-law function \( R_{\mathrm{max}} = R_0(N/N_0)^{\alpha_\mathrm{c}} \), where \textit{N} is the number of stars in the cluster and the power-law slope $\alpha_\mathrm{c} = 1/3$. The age distribution of protostars in Class 0, I, ``flat-spectrum'', II and III stages follows that of the Perseus low-mass star-forming cluster \citep[]{2009ApJS..181..321E,2014ApJ...787L..18S}. The model applies the Chabrier initial mass function (IMF) \citep{2003PASP..115..763C} for young clusters and disks. The outflow position angles are chosen randomly from 0$^{\circ}$ to 180$^{\circ}$, as well as the distance from the protostar to the outflow lobe with the maximum separation equal to \( 2 \times 10^4\) AU. The molecular outflow emission is assigned based on a scaling relation of the observed outflow emission from single low-mass protostars in the nearby low-mass star-forming regions NGC 1333 and Serpens Main and their modeled envelope masses. However, the emission is assigned only to Class 0 and I protostars, because ``flat-spectrum'', Class II and III objects only produce negligible molecular outflows \citep{2007prpl.conf..245A}. The cluster-in-a-box model focuses on the \( 7_0-6_0~\mathrm{A}^+\) methanol line at 338.409 GHz. 

    The cluster model did not include the contribution from high-mass sources, neither in the form of their outflows nor their hot cores. Nevertheless, a proof-of-concept study showed that the model reproduces the extended emission from a high-mass star-forming region to within a factor of two without tweaking the input parameters, suggesting that low-mass outflows account for $\sim$50\% of the total cluster emission. These results indicate that such a toy model can be used to constrain parameters of star-forming clusters and decipher the contribution from their components, i.e., molecular outflows and hot cores, and reproduce their morphologies.

\subsection{Galaxy-in-a-box}
\label{giab}

    New telescope facilities, particularly ALMA, are now routinely observing molecular emission at high redshift \citep[e.g., out to $z\gtrsim6$,][]{2017ApJ...842L..15S}. One possibility for understanding the origin of this emission is to use Galactic star-forming clusters as templates of emission. This approach would consist first of scaling Galactic observations to cover entire galaxies, and then comparing these scalings with actual observations of local galaxies. Next, the scalings would be extrapolated to the high-redshift ($z \gtrsim 1$) regime, where they can be compared to observations. Practically, the approach would consist of first creating a cluster model (Sect. \ref{ciab}), then populating a galaxy with these model clusters, thereby going from a cluster-in-a-box model to a galaxy-in-a-box model. This model consists of (i) a template (spiral) galaxy with molecular cloud spatial, age and mass distributions, and (ii) template stellar clusters with assigned outflow emission based on the cluster-in-a-box model. In this manner, emission from an entire galaxy may be simulated, with the advantage that the model only depends on a few input parameters. 
    
    Our knowledge about astrochemistry and star-formation primarily comes from observations of the Milky Way \citep[e.g.,][]{2009ARA&A..47..427H}. Thus, when first going to the extragalactic regime, the goal is to use the knowledge from the Milky Way together with a similar galaxy that could provide the pivotal information on its spatial structure. Furthermore, the galaxy should be nearby, well-studied, and ideally face-on, such that line-of-sight effects are minimized. One example of such a galaxy is the grand-design spiral ``Whirlpool Galaxy'', M51. Besides the spiral structure, M51 has an apparent size of 24 kpc \citep{2003AJ....125..525J}, which is roughly comparable to the estimated size of the Galactic disk $\gtrsim$ 30 kpc \citep{2016ARA&A..54..529B}. It is nearby \citep[$D\sim7.6$ Mpc;][]{2002ApJ...577...31C} and almost face-on \citep[$i\sim22^\circ$;][]{2014ApJ...784....4C}, making it an object of numerous studies, e.g.,  the Plateau de Bure Interferometer Arcsecond Whirlpool Survey \citep[PAWS;][]{2013ApJ...779...42S}. Therefore, in the following, we will base the template galaxy against observational data from M51.

    For the galaxy-in-a-box, we picked water as a default molecule to simulate galactic emission from. The reason for it is that from the 30\% of molecular species observed in the Milky Way, which were also detected in external galaxies \citep{2021arXiv210913848M}, water stands out as a ubiquitous star formation tracer in the Milky Way with emission dominated by molecular outflows and is readily observed towards high-$z$ galaxies \citep[e.g.,][]{2016A&A...595A..80Y,2017A&A...608A.144Y,2019ApJ...880...92J,2021A&A...648A..24V}. For the purpose of this work, we focused on the emission of the para-H$_2$O \(2_{02} - 1_{11}\) line at 987.927 GHz.

    In addition to the change of the molecular species used for obtaining the mass-intensity relation, the cluster model underwent a few upgrades while being adapted to the galactic model. One of the major changes is the spatial configuration defined in the cluster model. At a distance of $\geqslant$7.6 Mpc, the structure of individual clusters is practically unresolvable (1$''$ corresponds to $\sim$~40 pc). Therefore, the spatial component for the galactic model was discarded. Moreover, we used a novel distribution of protostellar ages following \cite{2018A&A...618A.158K}. We describe all of the relevant changes and upgrades motivated by scaling up the cluster model in greater detail in the following paragraphs. At first, we describe the spatial distribution applied in the galaxy model (Sect. \ref{spat}), then we define the molecular cloud mass distribution (Sect. \ref{mcmmass}), and from here, we go to the age distribution (Sect. \ref{age}).
    
\subsubsection{Spatial distribution}
\label{spat}

    \begin{figure}[t]
      \resizebox{\hsize}{!}{\includegraphics{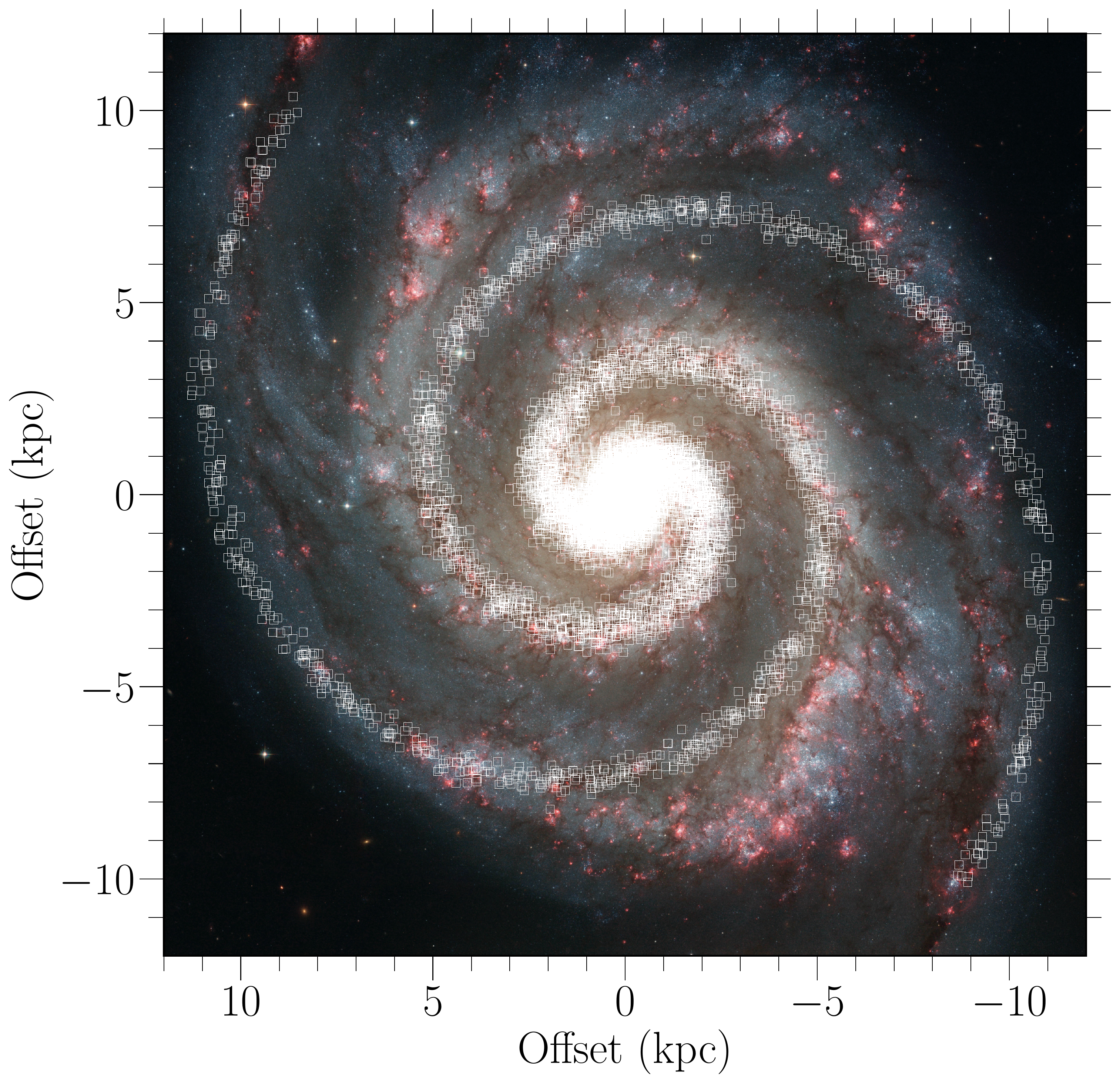}}
      \caption{Modelled two-part spatial configuration used in the galaxy-in-a-box model mapped onto the NASA's Hubble Space Telescope (credit: NASA, ESA, S. Beckwith (STScI), and the Hubble Heritage Team (STScI/AURA)) image of M51. The M51 image was scaled to fit within the spatial size settings used in the model. The white squares represent the location of stellar clusters along the spiral arms.}
    \label{m51}
    \end{figure}
    
    The spatial distribution of GMCs, in which young clusters form, in the galaxy-in-a-box model follows \cite{2009MNRAS.397..164R}:
    \begin{equation}
    \label{eq-gal}
        r(\phi) = \frac{A}{\mathrm{log}\, (B\, \mathrm{tan}\, \frac{\phi }{2N_\mathrm{S}})}
    \end{equation}
    where $A$ is a scale parameter for the entire structure, while $B$ and $N_\mathrm{S}$ determine the spiral pitch. This formula assumes that all galaxies have ``bars'' hidden within a bulge. Increasing the $N$ value results in tighter winding and $B$ in greater arm sweep and smaller bar/bulge. To emulate M51 we adopted the following values: $A = 8.0$, $B = 1.0$, and $N_\mathrm{S} = 8.26$. To obtain long spiral arms, wrapping around each other, we chose an angle coverage, $\phi$, of 500 degrees. We also introduced a direct scaling parameter $S = 1.5$ to shift spiral arms closer together, towards the galaxy center, without altering their spatial setups. This is especially useful to simulate a central bulge within a galaxy. The parameter is designed to be added at the end of Eq. \ref{eq-gal}.  The values were chosen to fit a galaxy with a $\sim23$ kpc diameter, which is roughly equivalent to the estimates of the M51 spatial size \citep[e.g.,][]{2003AJ....125..525J}. Figure \ref{m51} illustrates the quality of our fit.

    We built our radial distribution of stellar clusters by utilizing an exponential decline of stellar surface density, $\Sigma_{\mathrm{star}}$, with radius, $R$, in the following way:
    \begin{equation}
        \Sigma_{\mathrm{star}} = \mathrm{exp}(-R/h_R)
    \end{equation}
    where $h_R$ is a characteristic scale-length. Here, the exponential radial distribution corresponds to a probability density function for the location of stellar clusters along the spiral arms, which are then randomly located according to this function. We follow \cite{2017A&A...605A..18C} and use $h_R = 2.38~\mathrm{pc}$ value in this study. 
    
    The density distribution of stars in M51 resembles a skewed normal distribution \citep{2009A&A...494...81S}. Therefore, the model initially assigns a given stellar cluster a randomly generated location along the spiral arm, and then a random position along the cross section of the spiral arm given by the skewed normal distribution. Studies show \citep[e.g.,][]{2007A&A...471..765B,2015A&A...576A..33H} that the gas and dust density in galaxies typically decrease as a function of the radius from the center. Along with the stationary density wave predicting an age gradient across the arms, this decrease implies that star formation activity preferentially occurs in a narrowing band of the spiral arms. To simulate this effect, the standard deviation associated with the skewed normal distribution is scaled as a function of the distance from the center:
    \begin{equation}
        \sigma = (2+0.5r)^{-1}\ .
    \end{equation}
    This $\sigma$ value was arbitrarily chosen based on a qualitative good fit with observations of star-forming regions in M51 \citep{2011ApJS..193...19K}.

\subsubsection{Molecular cloud mass distribution}
\label{mcmmass}

    In the galaxy-in-a-box model, the initial number of GMCs is specified and then each GMC is randomly assigned a mass following the molecular cloud mass distribution. The latter is described by the molecular cloud mass probability density function (PDF):
    \begin{equation}
    \label{mcm}
        \dfrac{\mathrm{d}N}{\mathrm{d}M} \propto M^\alpha\ .
    \end{equation}
    We adopt a value of the slope, $\alpha=-1.64$ following \cite{2010ApJ...723..492R}. This value is in a good agreement with other Galactic studies of the GMCs, clouds and clumps \citep[e.g.,][]{1987ApJ...319..730S,2014MNRAS.443.1555U}. However, this power-law slope was derived for molecular clouds with masses between $10^5~ \mathrm{M}_\odot$--$10^6~\mathrm{M}_\odot$. Therefore, we assume that lower masses follow a similar slope and so we can use this $\alpha$ value for our study, where we utilize this relation for the mass range $10^4~\mathrm{M}_\odot$--$10^6~\mathrm{M}_\odot$. Estimates of extragalactic $\alpha$ show that this value probably is not constant among galaxies, and report variations reaching $\alpha \sim -3.0$, and estimate the average $\alpha \sim -2.0$ \citep[e.g.,][]{2005PASP..117.1403R, 2018MNRAS.477.5139G, 2020ApJ...893..135M}. We will evaluate the impact of different $\alpha$ values on the model in Sect. \ref{mcmres}.
    
    Subsequently, we use the mass distribution obtained with Eq.~\ref{mcm} to calculate the size of each molecular cloud. Here, we follow the recent estimate of the mass-size relation for Galactic GMCs from \cite{2020ApJ...898....3L}:
    \begin{equation}
    \label{massspec}
        R = 3.3 \times 10^{-3}~{\rm pc}\ \left( \frac{M}{M_\odot} \right)^{0.51}\ .
    \end{equation}
    
    To account for the fact that not all of the molecular cloud mass is converted to stellar mass, we assign a star formation efficiency, $\varepsilon_{\mathrm{SF}}$, to determine the total mass of the stellar population from the molecular cloud mass. In the model we apply $\varepsilon_{\mathrm{SF}} \sim 10\%$ for embedded clusters following \cite{2010ApJ...724..687L}.
    
    \begin{figure}[t]
    \resizebox{\hsize}{!}{\includegraphics{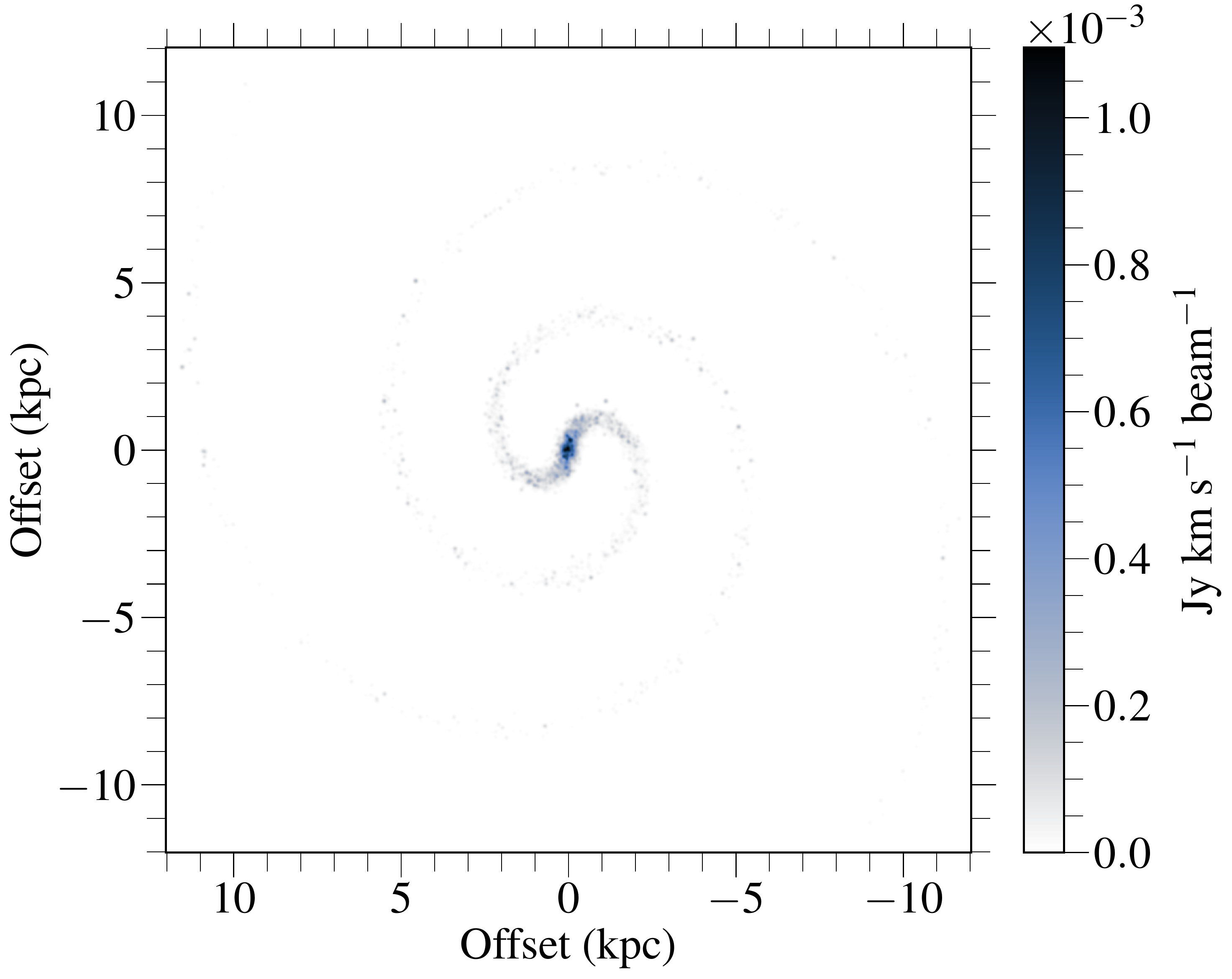}}
    \caption{Example integrated intensity map of the template galaxy with the standard setup.}
    \label{gal-temp}
    \end{figure}

\subsubsection{Age distribution}
\label{age}

    The characteristic time scale associated with star-forming regions is the free-fall time scale, $t_{\mathrm{ff}}$:
    \begin{equation}
        t_{\mathrm{ff}} = \sqrt{\frac{3\pi}{32G\rho}}
    \end{equation}
    where $\rho$ is the density of the cluster calculated as the total mass of the progenitor molecular cloud divided by the volume of the cloud. The free-fall time reflects the time required for a medium with negligible pressure support to gravitationally collapse. Here, we utilize this time scale to determine a lifetime of the clusters. However, not all of the molecular reservoir will undergo gravitational collapse. Recent studies find that $\varepsilon_{\mathrm{SF}}$  per $t_{\mathrm{ff}}$ remains constant among different molecular clouds \citep[e.g.,][]{2021ApJ...912L..19P}. To account for this inefficiency and its influence on the efficiency of $t_{\mathrm{ff}}$, we impose a scaling factor, $\tau_{\mathrm{ff}}^{\mathrm{sc}}$. In this study, we set the standard value of this factor to be 1. We also assume a constant free-fall time for the entire cluster.
    
    To assign a random age to the cluster we scale $t_{\mathrm{ff}}$ with the chosen $\tau_{\mathrm{ff}}^{\mathrm{sc}}$, and subsequently choose random values ranging between 0 (newly formed) and 1 (completely collapsed).
    The assigned ages are used to calculate the star formation rate, given by:
    \begin{equation}
        \lambda_{\mathrm{SF}} = \dfrac{N(t)}{t}
    \end{equation}
    where $N(t)$ is the number of stars at time $t$, which is the current age of the cluster calculated from the free-fall time. Here, we make an assumption that $\lambda_{\mathrm{SF}}$ is constant for the entire cluster.
    
    To assign the ages to protostars and determine their distributions within clusters, we follow \cite{2018A&A...618A.158K} and adopt a novel age distribution module. We start with the assumption that protostellar evolution is sequential, i.e., it begins at Class 0 and then goes through Class I, ``flat-spectrum'', Class II,  and ends at Class III. Then, with the constant star-formation rate and protostellar half-lives, sequential decay is applied. This decay, associated with protostars going through the evolutionary stages, is characterized by the ``decay'' constant $\lambda_{\mathrm{D}}$, where D represents the protostellar class. Values of $\lambda_{\mathrm{D}}$ for each evolutionary stage are estimated based on the observations of seven Galactic clouds \citep[for further details, see][]{2018A&A...618A.158K}. With this, we calculate the fractional population of stars in each evolutionary class for all galactic clusters.

     \begin{table*}[t]
    \caption{Overview of the most important global parameters in the galaxy-in-a-box model.}             
    \label{tab:stand}      
    \centering          
    \begin{tabular}{c| l c c c}
    \hline\hline       
    Category & Parameter & Description & Standard value & Ref.\\ 
    \hline
    \multirow{6}{*}{Star formation}
    & $x$                                     & \multicolumn{1}{l}{\begin{tabular}{@{}l@{}}power-law slope for the high-mass end of IMF, i.e,\\ for stars with masses $> 1 M_\odot$; $0 \rightarrow $ standard, $x = -2.3$;\\ $1 \rightarrow $ top-heavy, $x = -1.3$; $2 \rightarrow $ bottom-heavy, $x = -3.3$\end{tabular}} & 0 & 1\\  
    & $\varepsilon_\mathrm{SF}$               & \multicolumn{1}{l}{\begin{tabular}{@{}l@{}}star formation efficiency\end{tabular}} & $10\%$ & 2    \\
    & $\tau^\mathrm{sc}_\mathrm{ff}$    & \multicolumn{1}{l}{\begin{tabular}{@{}l@{}}free-fall time scaling factor\end{tabular}} & 1 & ... \\
    & $\alpha$                                & \multicolumn{1}{l}{\begin{tabular}{@{}l@{}}power-law slope of molecular cloud mass distribution\end{tabular}} & $-1.64$ & 3 \\
    & $N_\mathrm{CL}$                         & \multicolumn{1}{l}{\begin{tabular}{@{}l@{}}number of simulated clusters\end{tabular}} & $10^4$ & ... \\
    & $M_\mathrm{GMC}$                        & \multicolumn{1}{l}{\begin{tabular}{@{}l@{}}minimum and maximum masses of progenitor giant\\ molecular clouds  \end{tabular}}& \multicolumn{1}{c}{\begin{tabular}{@{}c@{}} $ 10^4 M_\odot \leqslant M_\mathrm{GMC} \leqslant 10^6 M_\odot$\end{tabular}} & ...  \\
    \hline
    \multirow{6}{*}{Morphology}
    & $A$                                     & \multicolumn{1}{l}{\begin{tabular}{@{}l@{}}galactic scaling factor\end{tabular}} & 8.0 & ... \\
    & $B$                                     &  \multicolumn{1}{l}{\begin{tabular}{@{}l@{}}galactic arms sweep\end{tabular}} & 1.0 & ... \\
    & $N_\mathrm{S}$                          & \multicolumn{1}{l}{\begin{tabular}{@{}l@{}}spiral winding number\end{tabular}} & 8.26 & ... \\
    & $\phi$                                  & \multicolumn{1}{l}{\begin{tabular}{@{}l@{}}angular coverage of spiral arms \end{tabular}} & $500^\circ$ & ... \\
    & $R_g$                                   & \multicolumn{1}{l}{\begin{tabular}{@{}l@{}}galactocentric radius\end{tabular}} & 12 kpc & 4\\
    & $h_R$                                   & \multicolumn{1}{l}{\begin{tabular}{@{}l@{}}characteristic scale-length\end{tabular}} & 2.38 pc & 5\\
    \hline
    \multirow{4}{*}{Observational}
    & $D$                                     & \multicolumn{1}{l}{\begin{tabular}{@{}l@{}}distance to the galaxy\end{tabular}} & 7.6 Mpc & 6\\
    & $\theta$                                & \multicolumn{1}{l}{\begin{tabular}{@{}l@{}}beam size convolved with galaxy image \end{tabular}} & 2\farcs55 & ... \\
    & $p_\mathrm{size}$                       & \multicolumn{1}{l}{\begin{tabular}{@{}l@{}} pixel size \end{tabular}} &  0\farcs51 & ... \\
    & dim                                     & \multicolumn{1}{l}{\begin{tabular}{@{}l@{}} image size in pixels \end{tabular}} &  1280 x 1280 & ... \\
    \hline                  
    \end{tabular}
    \tablebib{(1)~\citet{2003PASP..115..763C};
(2) \citet{2003ARA&A..41...57L}; (3) \citet{2010ApJ...723..492R}; (4) \citet{2002ApJ...577...31C};
(5) \citet{2017A&A...605A..18C}; (6) \citet{2003AJ....125..525J}.
}
\end{table*}

\subsection{Water Emission Database}
\label{wed}

    Our model relies on archival water observations. Thus, as a part of this project, we created the Water Emission Database (WED). The main goal of creating this database is to gather all of the available water data, from both ground-based observatories and the \textit{Herschel} Space Observatory, in one place and make it publicly available. This way, the data serves the scientific community. The database is stored and maintained using the MySQL Database Service. However, access to the data is granted through regularly updated ASCII and CSV files available online and is independent of the database driver for safety measures.

    Data from many Galactic surveys and observational projects are included in WED, e.g., Water In Star-forming regions with \textit{Herschel} \citep[WISH;][]{2011PASP..123..138V}, the William Herschel Line Legacy Survey \citep[WILL;][]{2017A&A...600A..99M}, Dust, Ice and Gas in Time \citep[DIGIT;][]{2013ApJ...770..123G}. Ultimately the database will also include extragalactic observations of water emission. The values that we store are particularly useful for this study. For example, we focused on water fluxes and parameters describing source properties. This means that we do not only store the values from specific studies, but we also keep a unified system of parameters important to characterize the sources. Currently, WED covers 79 observed water transitions up to the para-H$_2$O \( 9_{19} - 8_{08} \) transition at 5280.73 GHz (56.77 $\mu$m). Emitting sources at these transitions include the whole range of Galactic protostellar sources, with the majority of low-mass protostars. 

    \begin{table}[t]
    \caption{Description of WED table columns}
    \label{tab:wed} 
    \centering  
    \begin{tabular}{l l }
    \hline\hline  
    \noalign{\smallskip}
    Column & Description \\
    \noalign{\smallskip}
    \hline 
       obs\_id  & Ordinal number of the input   \\
       object  & Name of the object  \\
       obj\_type$^\mathrm{a}$  & Emitting object type \\
       ra\_2000  & RA (J2000)  \\
       dec\_2000  & Dec (J2000)  \\
       transition  & Observed water transition  \\
       freq$^\mathrm{b}$ & Rest frequency of the observed transition  \\
       telescope  & Name of the telescope used in the observations \\
       instrument$^\mathrm{c}$  & Instrument used in the observations \\
       obs\_res  & Resolution (\arcsec)   \\
       distance  & Distance to the object (pc)    \\
       luminosity  & Bolometric luminosity ($L_\odot$)  \\
       tbol$^\mathrm{c}$  & Bolometric temperature (K) \\
       menv$^\mathrm{c}$  & Envelope mass ($M_\odot$) \\
       vlsr$^\mathrm{c}$  & Velocity (km s$^{-1}$)  \\
       flux  & Observed water flux \\
       flux\_err$^\mathrm{c}$  & Flux error  \\
       unit  &   \multicolumn{1}{l}{\begin{tabular}{@{}l@{}} Unit of the observed flux \\(K km s$^{-1}$; W cm$^{-2}$; W m$^{-2}$; erg s$^{-1}$cm$^{-2}$) \end{tabular}}\\
       ref$^\mathrm{d}$ & Reference to the flux measurement(s) \\
       extra  & Other relevant information\\
       
    \hline  
    \end{tabular}
    \tablefoot{(a) Object types currently in use: YSO - Young Stellar Object, IM - Intermediate-mass, LM - Low-mass, IR-q - IR-quiet, HM - high-mass, mIR-q - mIR-quiet, HMPO - high-mass protostellar object, HMC - hot molecular core, UCHII - ultra-compact HII region, C0 - Class 0, CI - Class I, CII - Class II, PS - possible pre-stellar core, PDR - photodissociation region. Classification is based on the source papers; (b) All of the frequencies to corresponding transitions are taken from the LAMDA database \citep{2005A&A...432..369S}; (c) When available; (d) If more than one flux measurement is available, then the most recent or commonly used one is provided with the references to the remaining ones being stored in this column.}
    \end{table}
 
    The database holds the data in tables arranged in 20 columns (see Table \ref{tab:wed}) and shares them in the form of CSV and ASCII files available online on the project website\footnote{\url{https://katarzynadutkowska.github.io/WED/}}. All of the files that are available for download are fully described and updated, whenever there is a change in the database. The galaxy-in-a-box model downloads the data directly from the website, which makes the access to the model completely independent from the restricted MySQL server. 
 
    For the purpose of this work, we use a very particular subset of WED. We chose the data for para-H$_2$O \(2_{02} - 1_{11}\) line at 987.927 GHz. This water line is among the brightest H$_2$O transitions observed toward Galactic star-forming regions. Furthermore, it is not a ground-state transition, and so it only mildly suffers from self-absorption even toward high-mass objects \citep{2013A&A...554A..83V}. Finally, this transition is routinely observed toward extragalactic and even high-$z$ objects \citep[e.g.,][]{2016A&A...595A..80Y, 2017A&A...608A.144Y, 2019ApJ...880...92J}. The data available in WED for this particular line cover the whole range of sources and therefore gives a broad overview of water emission. \cite{2016A&A...585A.103S} identified an intensity$-$envelope mass relation for this line, $\mathrm{log}L_\mathrm{H_2O}=(-2.91\pm0.10) + (1.19\pm0.05)\cdot\mathrm{log}M_\mathrm{env}$, which we also observe for the data used in this study (see Fig. \ref{menv-tdv}).
    As mentioned, the emission assignment utilizes the relationship between the line intensity and envelope mass. At first, Class 0 and Class I objects are assigned with a stellar mass sampled from the IMF. Then we subsequently convert the stellar masses to envelope masses by assuming the envelope mass corresponds to $3\times$ and $1.5\times$ stellar mass for Class 0 and I protostars, respectively \citep[e.g.,][and for a more in-depth discussion \citealt{2014offner}]{2010A&A...518L.102A}. Following this, by using the intensity$-$envelope mass relation, we assign outflow emission to these deeply embedded protostars. We build this relation for para-H$_2$O \(2_{02} - 1_{11}\) line data from the WISH and WILL samples. The observed intensities are distance-normalized to get a distance-independent measurement. To assess the goodness-of-fit of the correlation in our regression model, we examined its R-squared value, which, in this case, corresponds to 89\%, indicating a strong relationship between envelope mass and intensity. We derived the correlation to follow:
    \begin{equation*}
        \mathrm{log} I_\nu~(\mathrm{Jy~km~s}^{-1})= -6.42\pm0.08 + (1.06\pm0.04)\cdot \mathrm{log} M_\mathrm{env}(\mathrm{M}_\odot)\ ,
    \end{equation*}
    where the intensity is normalized to the distance of M51, i.e., 7.6 Mpc. From the above correlation we see that there is a near-proportionality between $I_{\nu}$ and $M_\mathrm{env}$, $I_\nu \propto M_\mathrm{env}$. 
    
    \begin{figure}[t]
    \resizebox{\hsize}{!}{\includegraphics{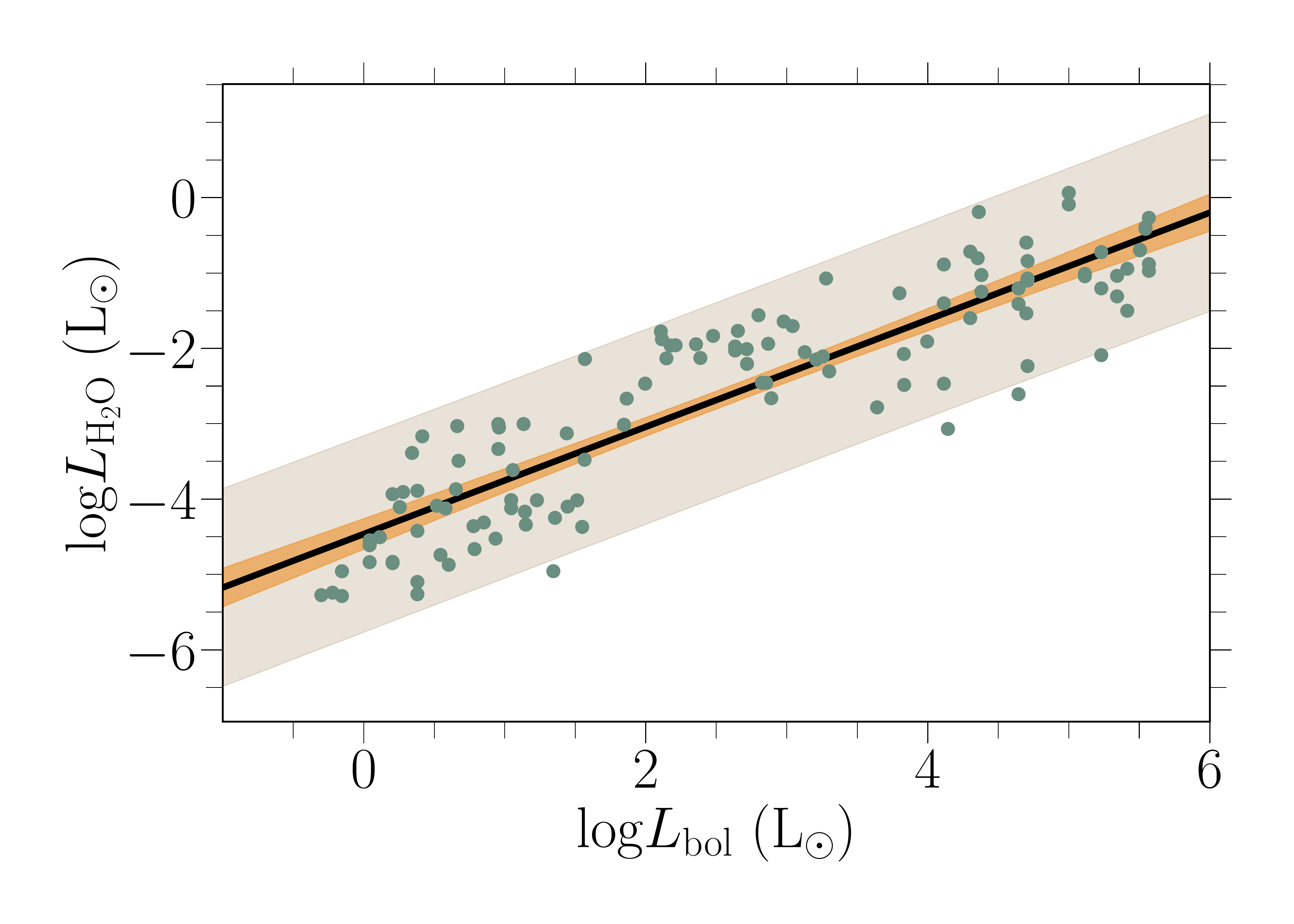}}
    \caption{Water line luminosity at 988 GHz vs. bolometric luminosity for objects from WED used in the simulations, with inclusion of few additional sources, that were excluded from the simulations due to lack of $M_\mathrm{env}$ data \citep{2015PhDT.......554S}. The solid black line shows the best-fit proportionality, the orange filled region corresponds to the 95\% confidence region of the correlation, and the shaded red region represents the region that contains 95\% of the measurements.}
    \label{lbol-lh2o}
    \end{figure}
    
    \begin{figure}[t]
    \resizebox{\hsize}{!}{\includegraphics{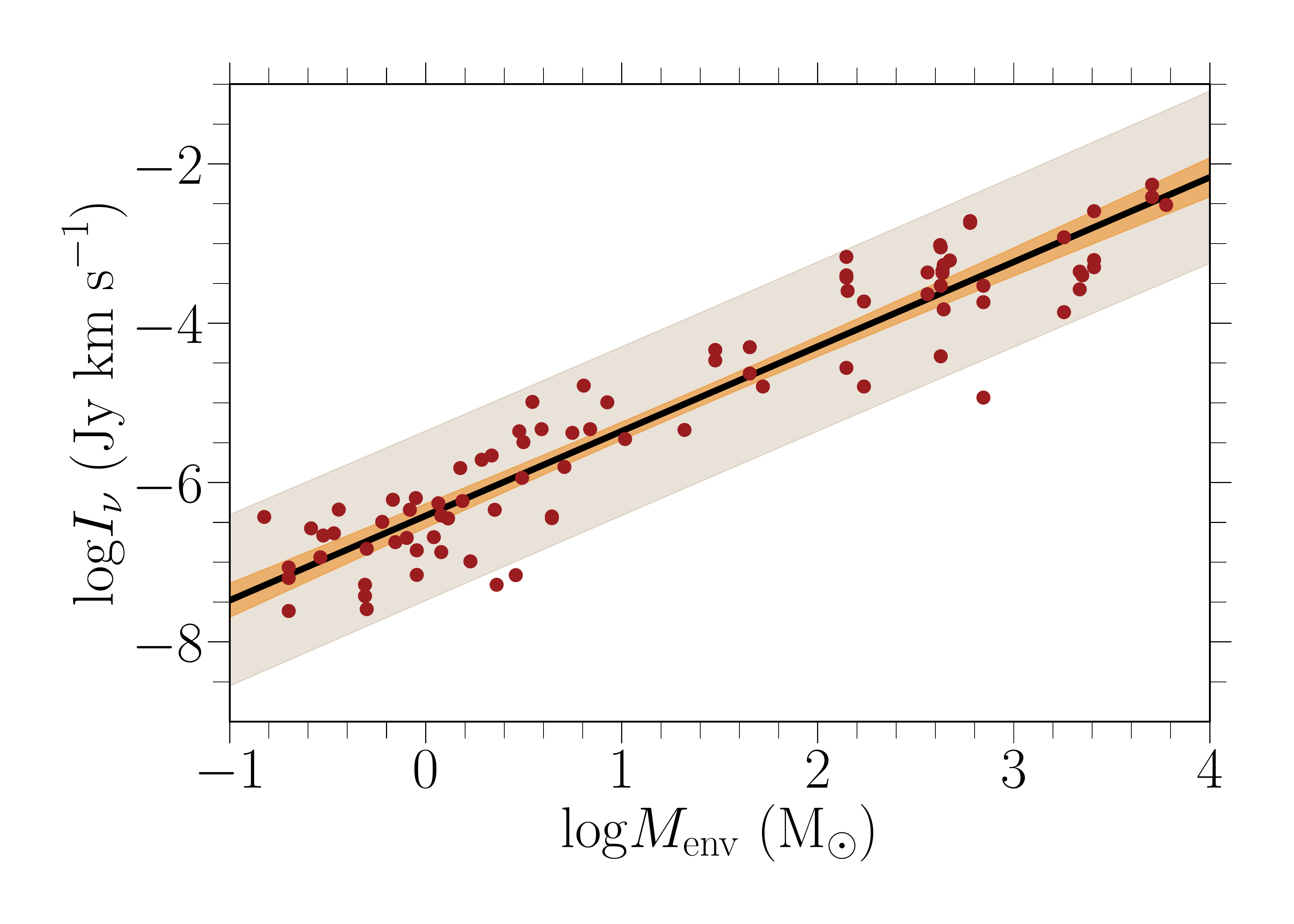}}
    \caption{Water emission at 998 GHz vs. envelope mass, $M_\mathrm{env}$, for objects from WED used in the simulations. Colors as in Fig. \ref{lbol-lh2o}.}
    \label{menv-tdv}
    \end{figure}

\section{Results}
\label{results}
    
     With the default galactic and star-formation parameters described in Sect. \ref{giab} and gathered in Table \ref{tab:stand}, we get an integrated intensity map of the desired molecular emission, as well as mass, total emitted emission and number of stars of each star-forming cluster within the simulated galaxy. An example integrated intensity map for the model with default parameters is presented in Fig. \ref{gal-temp}. With the chosen spatial setup, most of the emission comes from the inner-most parts of the galaxy, where the bulge is located and here individual clusters are not resolved with the applied beam size of 2\farcs55 (see Table \ref{tab:stand}). The farther from the bulge, the lower the emission and the easier it is to resolve clusters within spiral arms, although the surface brightness of course also decreases. 
     
     To explore the impact of the global star-formation parameters on the expected emission from clusters in a simulated galaxy as well as the galaxy itself, we conducted a parameter-space study. The changes in parameters were set with respect to the standard model configuration (Table \ref{tab:stand}). We focused on the variations caused by the most important global SF-related parameters, i.e., (i) $\alpha$, describing the slope of molecular cloud mass distribution, (ii) $\varepsilon_\mathrm{SF}$, the star-formation efficiency per free-fall time, (iii) $\tau^\mathrm{sc}_\mathrm{ff}$, the free-fall scaling parameters, and (iv) the power-law slope for the high-mass end of IMF. For each change in parameters, we run 10 simulations to derive the average of predicted emission, while for the standard setup we decided on 30 model runs to lower the variations in the derived values. The choice of running 10 simulations was motivated by cutting down on the computational time, and it is enough to show the variability in the model outcomes. We explored the cumulative impact of these parameters on the total galactic emission, radial profiles of the emission maps, and distributions of emitted flux by the galactic clusters. As will be shown below, these seem to be consistently skewed. Therefore, we chose median values as a measure of central tendency and explored the spread of these distributions with the interquartile range method (IQR or midspread), providing information on the middle 50\% of values with the median being in the center of the range.
    
    \begin{figure}[h!]
      \resizebox{\hsize}{!}{
      \subfloat{\includegraphics{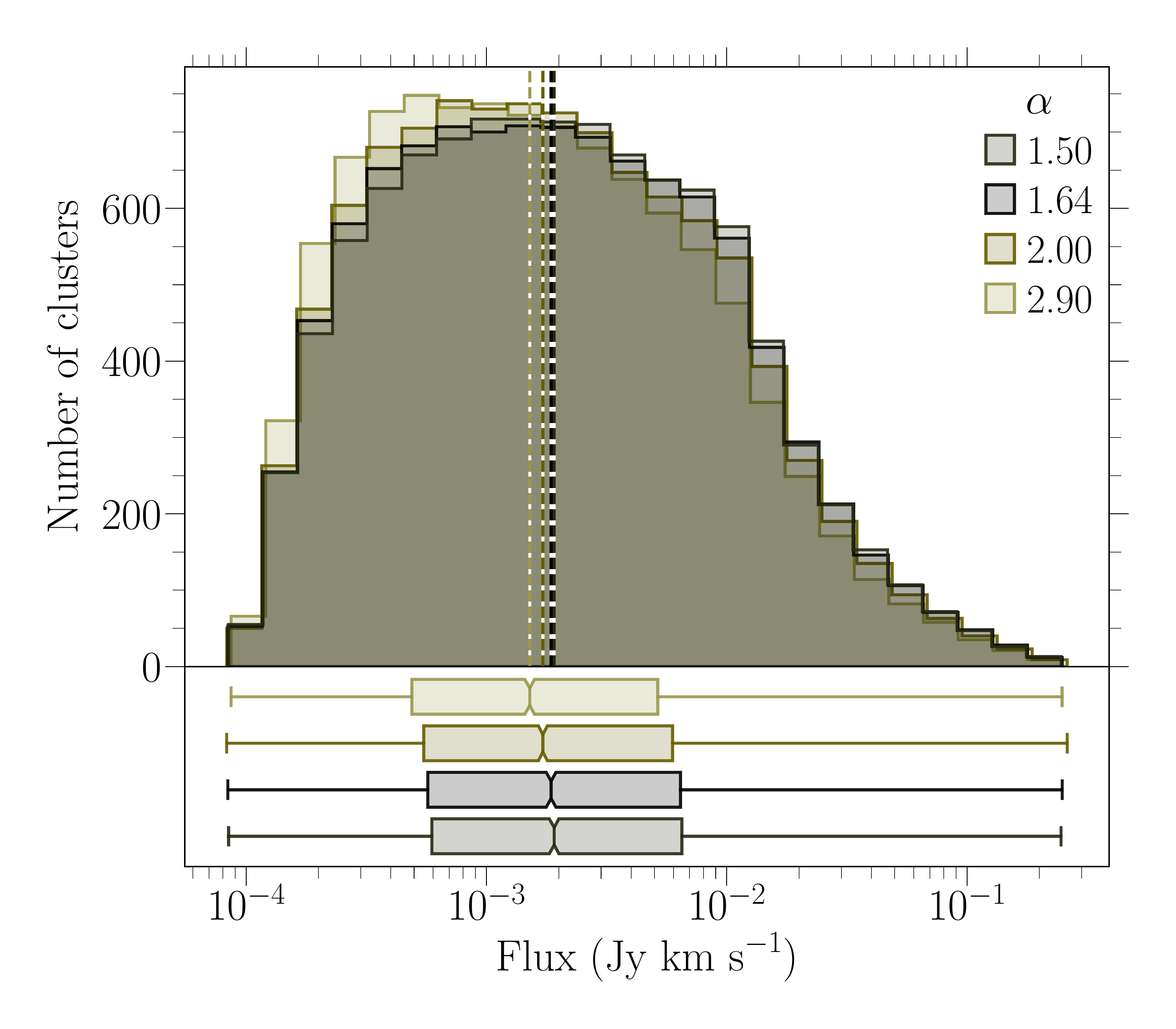}} 
      }\\[-2ex]
      \resizebox{\hsize}{!}
      {\subfloat{\includegraphics{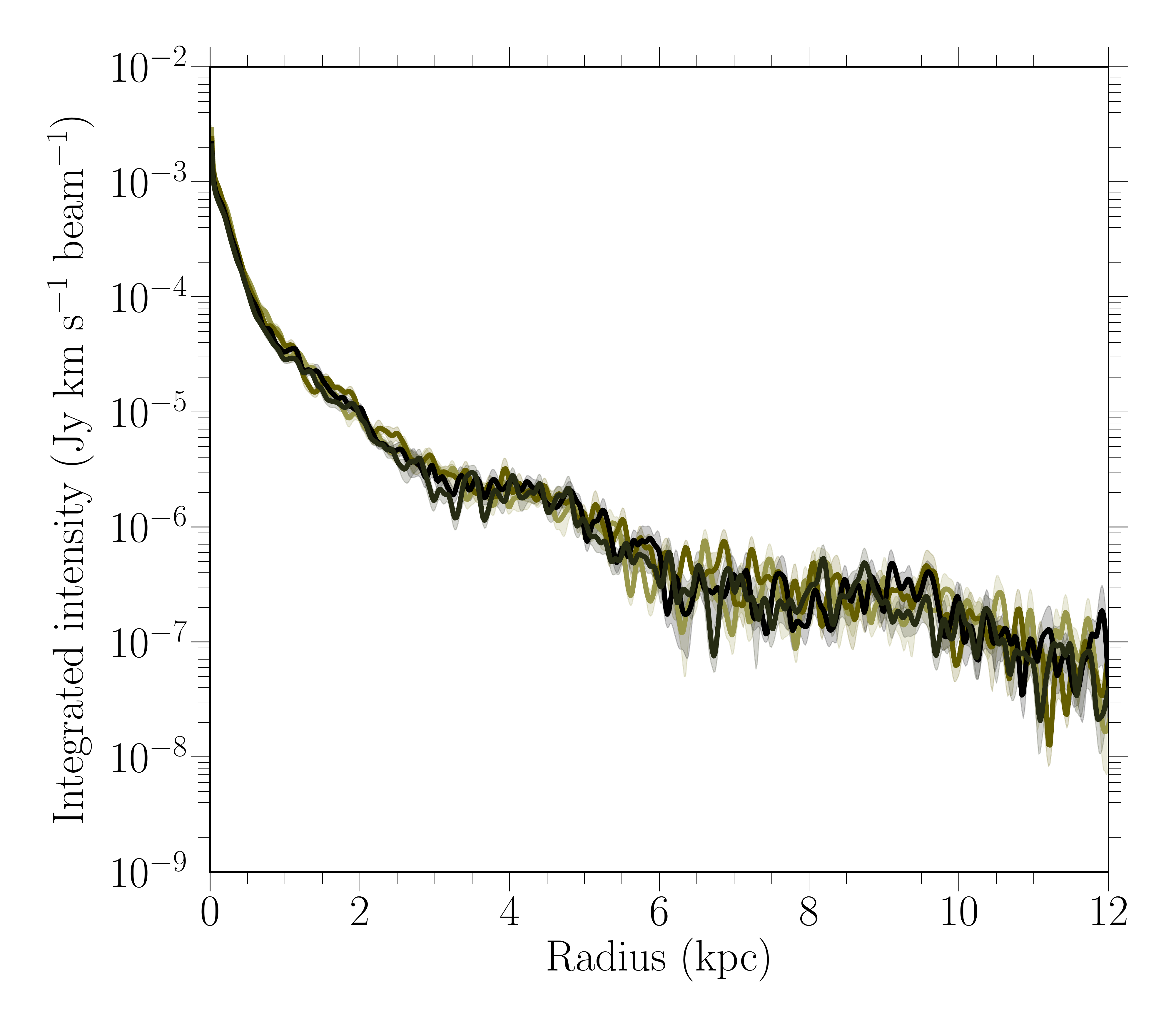}} 
      }
      \caption{\textbf{Top.} Distributions of cluster emission derived for simulations with different power-law slopes $\alpha$ of molecular cloud mass distributions. The vertical dashed lines correspond to the median flux of each distribution. In the bottom box plot the interquartile ranges are presented. The notches in the boxes indicate the 95\% confidence intervals around the median. The whiskers spread to the beginning (0\%) and the end (100\%) of the distributions. These are the mean distributions from a series of 10 simulations for each varying parameter.  \textbf{Bottom.} Radial profiles of emission from the galaxies of the corresponding $\alpha$ values. The radial profiles were calculated from the center of the galaxy all the way to its outskirts. The solid lines correspond to the mean profiles derived from 10 simulations, while the shaded regions represent the spread of the mean values based on their standard deviations.}\label{alpha}
    \end{figure}
    
\subsection{Molecular cloud mass distributions}
\label{mcmres}
    
    The standard value of $\alpha$ is set to $-1.64$ \citep{2010ApJ...723..492R}. Different studies \citep[e.g.,][]{1987ApJ...319..730S,2005PASP..117.1403R,2020ApJ...893..135M} report a spread in $\alpha$ depending on the studied regions, and following these studies we explore the change in expected emission for $\alpha = -1.5$, $-2$ and $-2.9$. The highest $\alpha$ follows the steepest index reported by \cite{2005PASP..117.1403R}. To investigate this impact we compared the distributions of flux emitted by the clusters and radial profiles of galactic emission.
    
    We observe no apparent variations in the expected emission caused by the change in $\alpha$. It is true both for the flux distributions as well as for the mean radial profiles Fig. \ref{alpha}). However, looking at the values obtained for the molecular cloud mass distribution (see Table \ref{tab:alpha}) we see a clear trend, indicating that with increasing $\alpha$, the median flux, the total galactic emission and interquartile range increase. This result is consistent with the nature of the corresponding mass distributions, as the steeper the slope the more emission comes from low-mass clusters, which in turn lowers the total observed emission.
    
    \begin{table}[t]
    \caption{Simulation results for different molecular cloud mass distributions}
    \label{tab:alpha} 
    \centering
    \begin{tabular}{c c c c}
    \hline\hline
    \noalign{\smallskip}
    $\alpha$ & 
    \multicolumn{1}{c}{\begin{tabular}{@{}c@{}} $\tilde{I}$\\$[\mathrm{Jy~km~s}^{-1}]$ \end{tabular}} & \multicolumn{1}{c}{\begin{tabular}{@{}c@{}} $I_\mathrm{tot}$\\$[\mathrm{Jy~km~s}^{-1}]$ \end{tabular}} & 
    \multicolumn{1}{c}{\begin{tabular}{@{}c@{}} IQR \\ $[\mathrm{Jy~km~s}^{-1}]$ \end{tabular}} \\ 
    \noalign{\smallskip}
    \hline
       -1.50  & $1.91\times10^{-3}$ & $7.04\times10^{1}$    & $5.91\times10^{-3}$\\ 
       -1.64  & $1.86\times10^{-3}$ & $7.02\times10^{1}$    & $5.85\times10^{-3}$ \\
       -2.00  & $1.72\times10^{-3}$ & $6.61\times10^{1}$    & $5.40\times10^{-3}$ \\
       -2.90  & $1.51\times10^{-3}$ & $5.91\times10^{1}$    & $4.68\times10^{-3}$ \\
    \hline
    \end{tabular}
    \tablefoot{Results from running 10 simulations per model configuration; $\alpha$ - power-law slope of the molecular cloud mass distribution, $(\tilde{I})$ - median flux, $(I_\mathrm{tot})$ - total galactic emission, IQR - midspread}
    \end{table}

\subsection{Initial mass function}
    
    In the model, we adopted three types of IMF based on the \cite{2003PASP..115..763C} IMF form for young clusters and disk stars. By introducing changes in the slope of the high-mass end of the IMF, $x$, which applies for stars with $M_\star > 1M_\odot$, we defined bottom- and top-heavy forms. With the standard value of $x = -2.3$, the slope for the bottom-heavy IMF is defined as $x-1$, while for the top-heavy it is $x+1$. This is a purely empirical parametrization, although it is in reasonable agreement with studies reporting $x$ values for bottom- and top-heavy IMF forms \citep[for a recent review, see][]{2020ARA&A..58..577S}.
    
    \begin{table}[t]
    \caption{Simulation results for different IMF configurations}     
    \label{tab:imf} 
    \centering    
    \begin{tabular}{c c c c}   
    \hline\hline  
    \noalign{\smallskip}
    IMF & 
    \multicolumn{1}{c}{\begin{tabular}{@{}c@{}} $\tilde{I}$\\$[\mathrm{Jy~km~s}^{-1}]$ \end{tabular}} & \multicolumn{1}{c}{\begin{tabular}{@{}c@{}} $I_\mathrm{tot}$\\$[\mathrm{Jy~km~s}^{-1}]$ \end{tabular}} & 
    \multicolumn{1}{c}{\begin{tabular}{@{}c@{}} IQR \\ $[\mathrm{Jy~km~s}^{-1}]$ \end{tabular}} \\
    \noalign{\smallskip}
    \hline 
       top-heavy    & $2.51\times10^{-3}$ & $8.57\times10^{1}$  & $7.87\times10^{-3}$ \\
       standard     & $1.86\times10^{-3}$ & $7.02\times10^{1}$  & $5.85\times10^{-3}$ \\
       bottom-heavy & $1.81\times10^{-3}$ & $6.90\times10^{1}$  & $5.73\times10^{-3}$ \\
    \hline
    \end{tabular}
    \tablefoot{Results from running 10 simulations per model configuration; IMF - form of the initial mass function, $(\tilde{I})$ - median flux, $(I_\mathrm{tot})$ - total galactic emission, IQR - midspread}
    \end{table}
    
\begin{figure}[t!]
      \resizebox{\hsize}{!}{
      \subfloat{\includegraphics{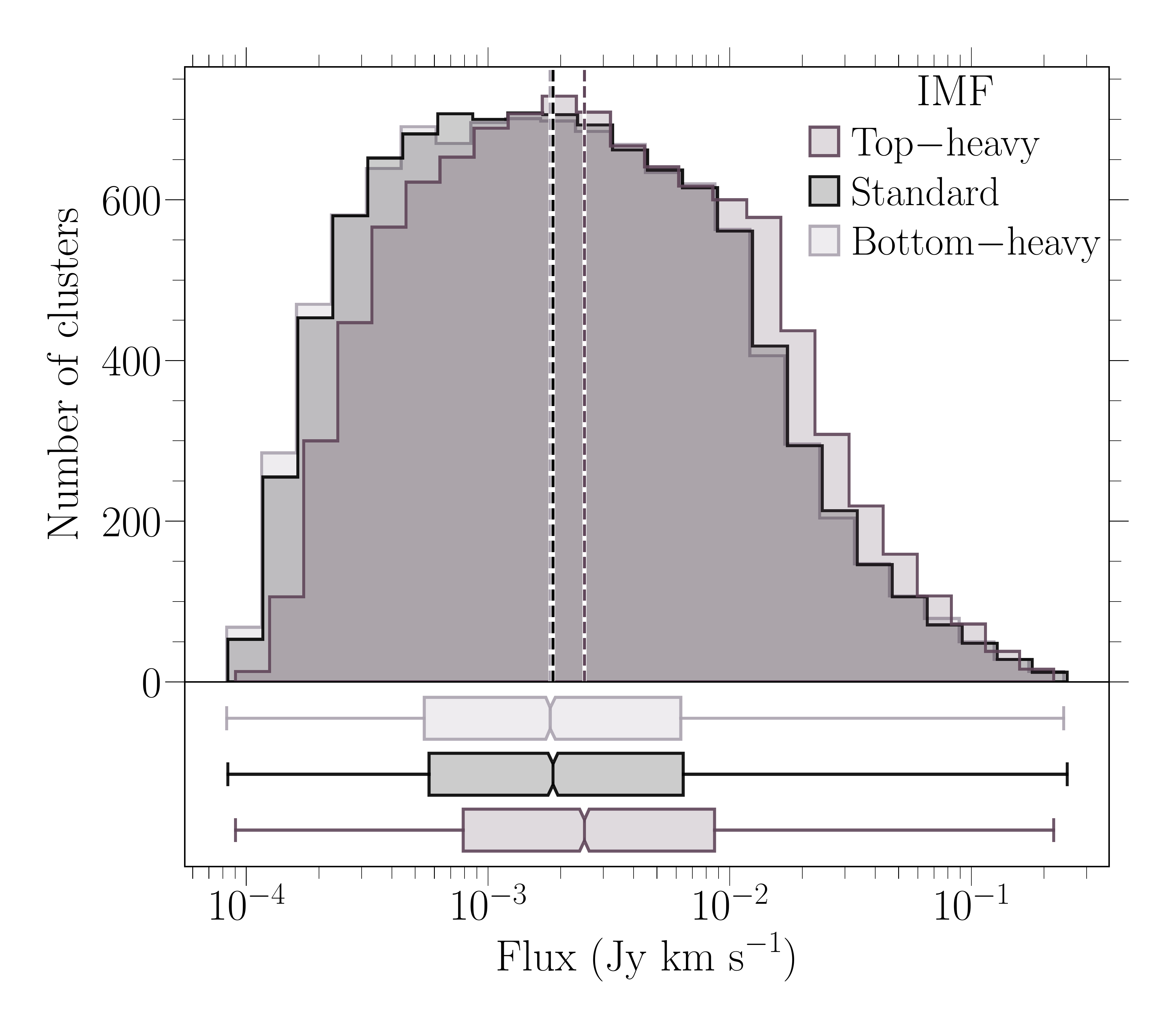}} 
      }\\[-2ex]
      \resizebox{\hsize}{!}
      {\subfloat{\includegraphics{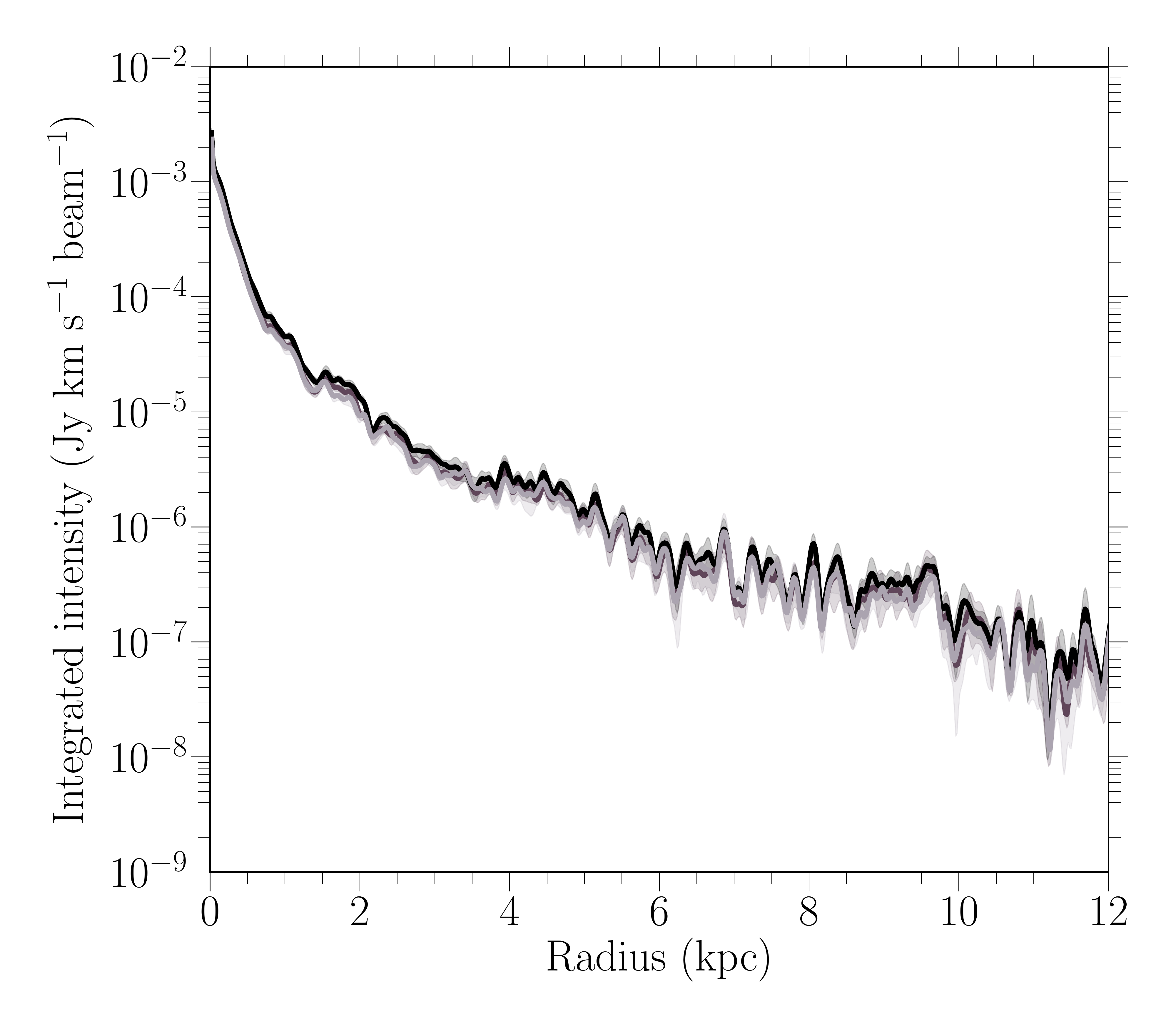}} 
      }
      \caption{As Fig. \ref{alpha} but for different IMF forms.} \label{imf}
    \end{figure}

    There is no apparent difference in examined values for any of the IMF types (see Table \ref{tab:imf}), although it is clear that our top-heavy IMF model tends to produce slightly more emission over the bottom-heavy one. We will discuss this further in Sect.\ref{discussion}. The lack of dominance of any IMF type is also true for the mean radial profiles of galaxies as depicted in Fig. \ref{imf}. Here, we see that neither around the inner part of spiral arms nor around their outer parts any of the considered IMF types take over the emission and the radial profiles are indistinguishable.
    
\subsection{Star-formation efficiencies}
\label{sfe-chap}
\begin{figure}[t!]
      \resizebox{\hsize}{!}{
      \subfloat{\includegraphics{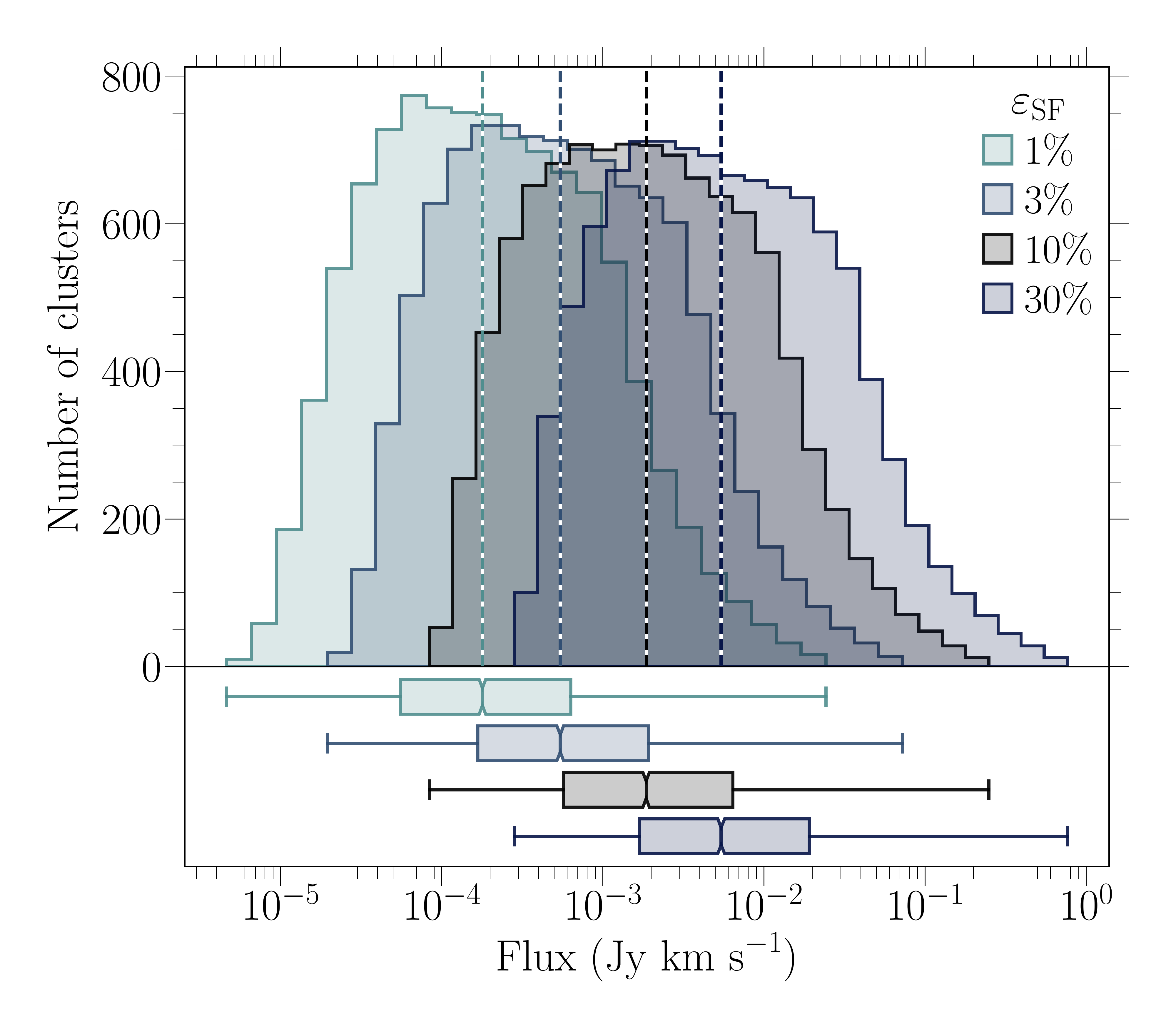}}
      }\\[-2ex]
      \resizebox{\hsize}{!}
      {\subfloat{\includegraphics{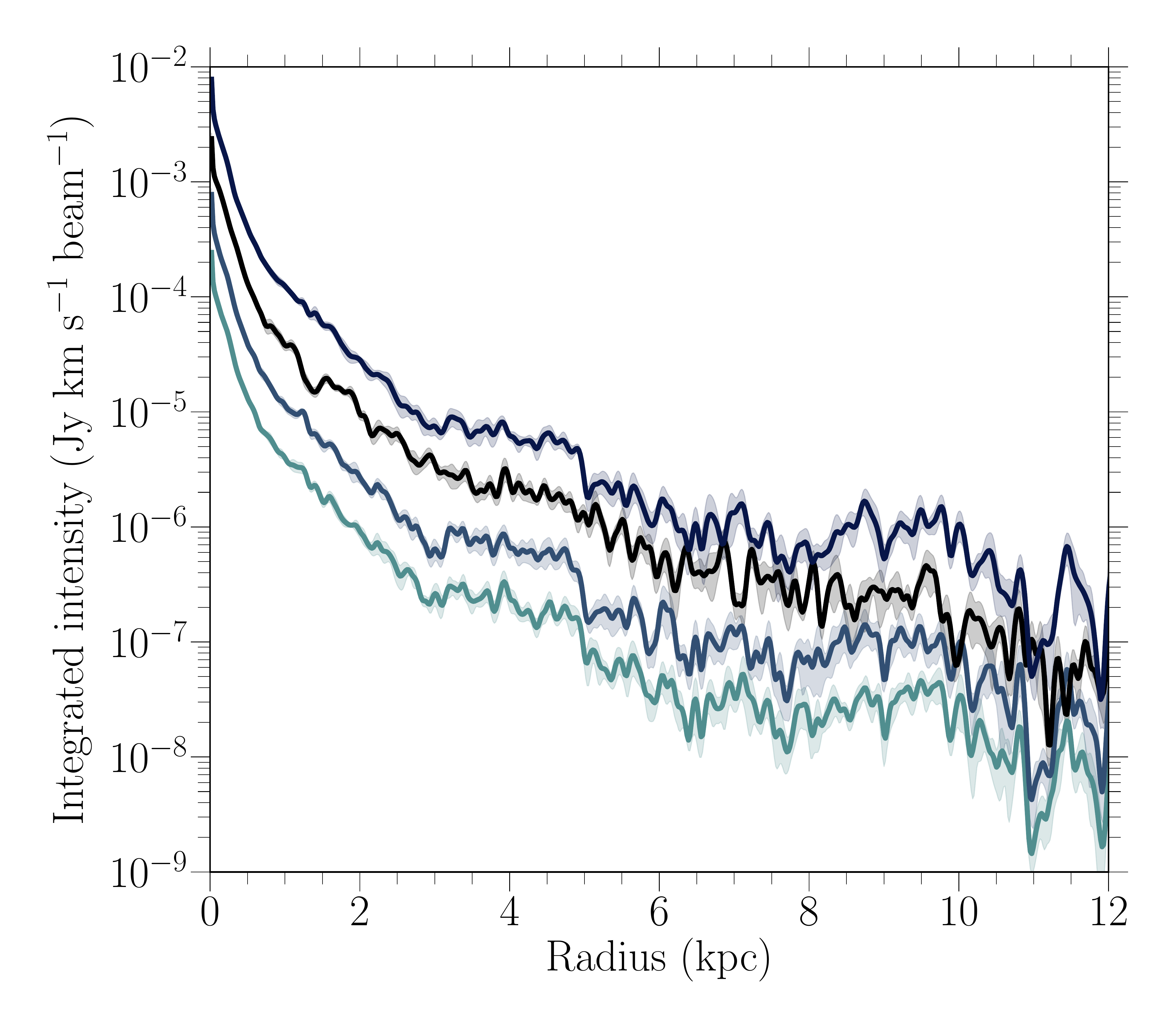}} 
      }
      \caption{As Fig. \ref{alpha} but for the varying $\varepsilon_\mathrm{SF}$.} \label{sfe}
    \end{figure}

    We probed the impact of $\varepsilon_\mathrm{SF}$ on emission outputs by varying its values from $1\%$ to $30\%$. The outputs vary strongly between different $\varepsilon_\mathrm{SF}$ values with a clear trend of increasing flux with $\varepsilon_\mathrm{SF}$ as seen in Fig. \ref{sfe}. The difference between the highest and lowest values roughly corresponds to one order of magnitude for all of the considered values. Moreover, we see that the shape of the distribution does not vary significantly across different $\varepsilon_\mathrm{SF}$ values, instead higher $\varepsilon_\mathrm{SF}$ merely translates distributions to higher flux values. This way, for the lowest $\varepsilon_\mathrm{SF}=1\%$ we derived the total galactic emission of $6.96~\mathrm{Jy~km~s}^{-1}$, while one order of magnitude higher $\varepsilon_\mathrm{SF}=10\%$ results in approximately one order of magnitude increase of the same parameter, giving  $7.02\times10^{1}~ \mathrm{Jy~km~s}^{-1}$. Besides the total galactic emission, $I_\mathrm{tot}$, this trend holds for the median fluxes, $\tilde{I}$, as well as for the midspreads, and it is clear that the multiplication of $\varepsilon_\mathrm{SF}$ on average corresponds to the same multiplication of flux (see Table \ref{tab:sfe}).
    
    \begin{table}[t]
    \caption{Simulation results for different star formation efficiencies} 
    \label{tab:sfe} 
    \centering     
    \begin{tabular}{c c c c} 
    \hline\hline     
    \noalign{\smallskip}
    $\varepsilon_\mathrm{SF}$ & 
    \multicolumn{1}{c}{\begin{tabular}{@{}c@{}} $\tilde{I}$\\$[\mathrm{Jy~km~s}^{-1}]$ \end{tabular}} & \multicolumn{1}{c}{\begin{tabular}{@{}c@{}} $I_\mathrm{tot}$\\$[\mathrm{Jy~km~s}^{-1}]$ \end{tabular}} & 
    \multicolumn{1}{c}{\begin{tabular}{@{}c@{}} IQR \\ $[\mathrm{Jy~km~s}^{-1}]$ \end{tabular}} \\
    \noalign{\smallskip}
    \hline                       
       $1\%$  & $1.79\times10^{-4}$  & 6.96 & $5.78\times10^{-4}$ \\
       $3\%$  & $5.45\times10^{-4}$  & $2.10\times10^{1}$    & $1.76\times10^{-3}$ \\
       $10\%$ & $1.86\times10^{-3}$  & $7.02\times10^{1}$    & $5.85\times10^{-3}$ \\
       $30\%$ & $5.43\times10^{-3}$  & $2.11\times10^{2}$    & $1.75\times10^{-2}$ \\
    \hline
    \end{tabular}
    \tablefoot{Results from running 10 simulations per model configuration; $\varepsilon_\mathrm{SF}$ - star formation efficiency, $(\tilde{I})$ - median flux, $(I_\mathrm{tot})$ - total galactic emission, IQR - midspread}
    \end{table}
    
    From mean radial profiles (see Fig. \ref{sfe}) it is also clear that the increase in the $\varepsilon_\mathrm{SF}$ value results in a subsequent increase of average emission from the galaxy. Here, the highest differences in intensities are also around one order of magnitude. Therefore, the higher the $\varepsilon_\mathrm{SF}$, the more emission comes from spiral arms at different points of the radius. Also, for $\varepsilon_\mathrm{SF}= 1\%$ and $\varepsilon_\mathrm{SF}= 3\%$, the drop in emission in the outermost parts of the galaxy results in higher variations and more significant drop of the observed emission.  
    
\subsection{Free-fall-time scaling}
    
    We studied the impact of the free-fall time in the form of $\tau_{\mathrm{ff}}^{\mathrm{sc}}$ by adopting values ranging from $\tau_{\mathrm{ff}}^{\mathrm{sc}}=0.5$ to $\tau_{\mathrm{ff}}^{\mathrm{sc}} = 5.0$. The scaling factor introduced in this study represents how many free-fall times it takes to form most of the stellar population in a single cluster and relates to the free-fall time efficiency as $\epsilon_{\mathrm{ff}} = 0.9\dfrac{M_*}{M_\mathrm{tot}}\dfrac{t_{\mathrm{ff}}}{t_\mathrm{form}}=0.9\dfrac{M_*}{M_\mathrm{tot}\tau_{\mathrm{ff}}^{\mathrm{sc}}}$ following \cite{2014rio}, where they estimated time required to form 90\% of stars in the cluster. Therefore, with this choice of the $\tau_{\mathrm{ff}}^{\mathrm{sc}}$ values, we evaluate the impact of the free-fall time efficiencies spreading over one order of magnitude, between $\epsilon_{\mathrm{ff}}\sim0.01 - 0.1$. 

    \begin{table}[t]
    \caption{Simulation results for different free-fall time scaling factors}
    \label{tab:eff} 
    \centering                         
    \begin{tabular}{c c c c}        
    \hline\hline        
    \noalign{\smallskip}
    $\tau^\mathrm{sc}_\mathrm{ff}$ & 
    \multicolumn{1}{c}{\begin{tabular}{@{}c@{}} $\tilde{I}$\\$[\mathrm{Jy~km~s}^{-1}]$ \end{tabular}} & \multicolumn{1}{c}{\begin{tabular}{@{}c@{}} $I_\mathrm{tot}$\\$[\mathrm{Jy~km~s}^{-1}]$ \end{tabular}} & 
    \multicolumn{1}{c}{\begin{tabular}{@{}c@{}} IQR \\ $[\mathrm{Jy~km~s}^{-1}]$ \end{tabular}} \\  
    \noalign{\smallskip}
    \hline
       0.5  & $3.36\times10^{-3}$  & $1.12\times10^{2}$    & $1.03\times10^{-2}$ \\
       1.0  & $1.86\times10^{-3}$  & $7.02\times10^{1}$    & $5.85\times10^{-3}$ \\
       3.0  & $6.06\times10^{-4}$  & $2.94\times10^{1}$    & $2.00\times10^{-3}$ \\
       5.0  & $3.64\times10^{-4}$  & $1.91\times10^{1}$    & $1.21\times10^{-3}$ \\
    \hline 
    \end{tabular}
    \tablefoot{Results from running 10 simulations per model configuration; $\tau^\mathrm{sc}_\mathrm{ff}$ - free-fall time scaling factor, $(\tilde{I})$ - median flux, $(I_\mathrm{tot})$ - total galactic emission, IQR - midspread}
    \end{table}
    
\begin{figure}[t!]
      \resizebox{\hsize}{!}{
      \subfloat{\includegraphics{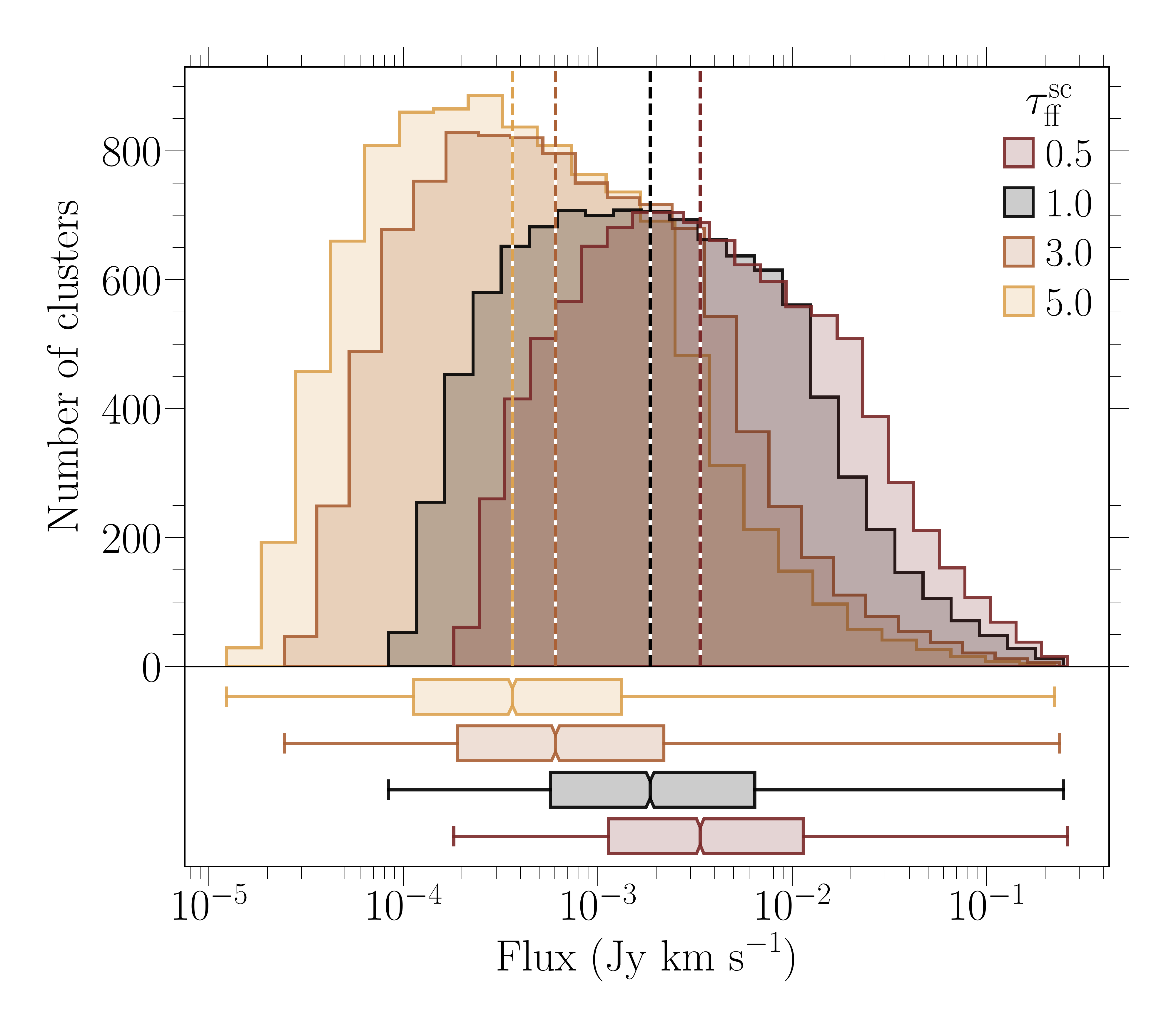}} 
      }\\[-2ex]
      \resizebox{\hsize}{!}
      {\subfloat{\includegraphics{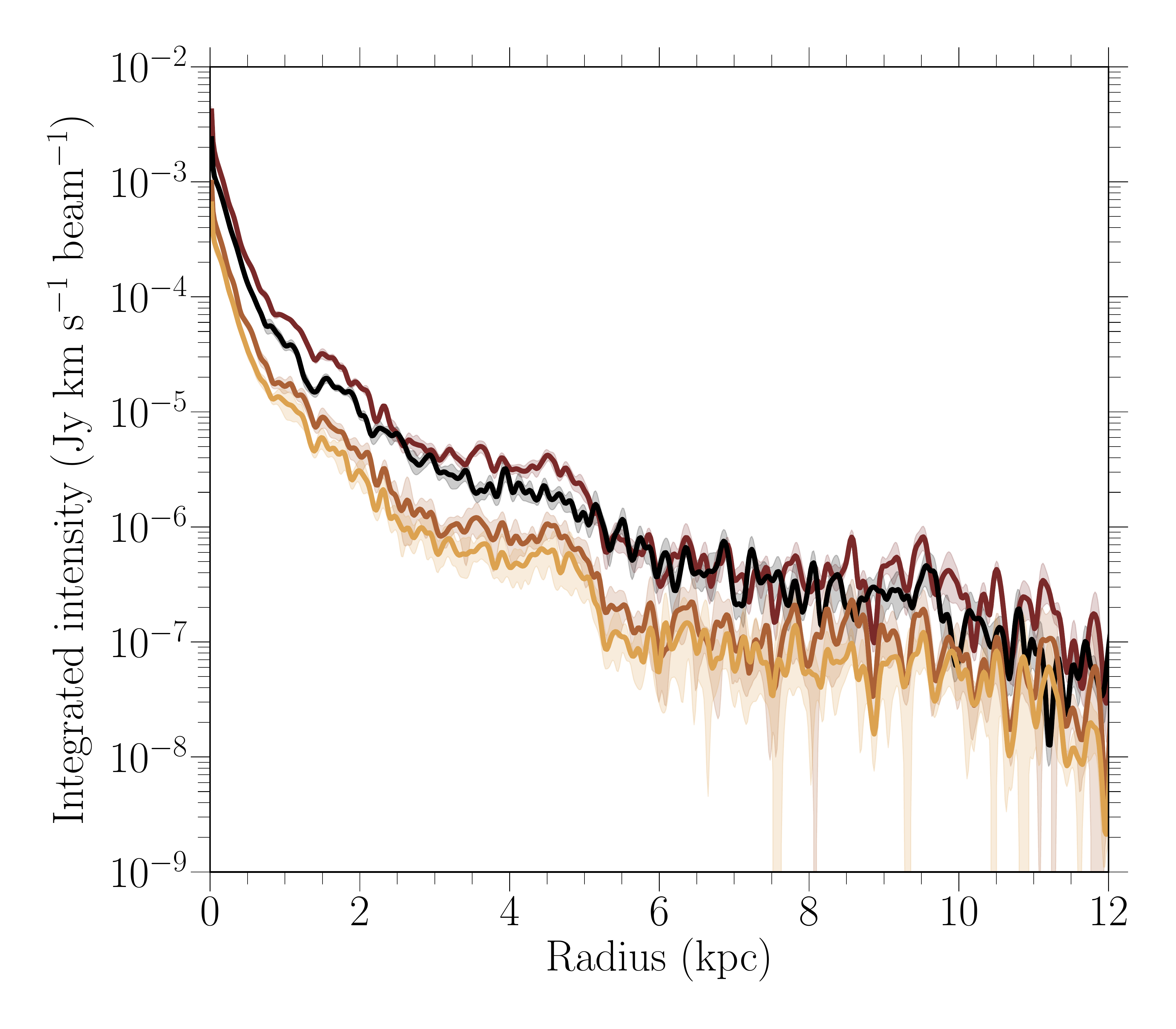}} 
      }
      \caption{As Fig. \ref{alpha} but for galaxies with different $\tau_{\mathrm{ff}}^{\mathrm{sc}}$.} \label{eff}
    \end{figure}

    We observe a very distinct relation between emitted flux and $\tau_{\mathrm{ff}}^{\mathrm{sc}}$ values, namely that with the decreasing $\tau_{\mathrm{ff}}^{\mathrm{sc}}$ the observed total flux increases. Moreover, the decreasing $\tau_{\mathrm{ff}}^{\mathrm{sc}}$ is associated with condensation of flux distributions, which get both narrower and flatter, and are shifted towards higher flux values (see Fig. \ref{eff}). The lowest $\tau_{\mathrm{ff}}^{\mathrm{sc}}$ results in the median flux value that is one order of magnitude higher than the one derived for the highest $\tau_{\mathrm{ff}}^{\mathrm{sc}}$ (see Table \ref{tab:eff}). Also, the beginnings of each distribution are shifted by one order of magnitude from $\sim10^{-5}$ to $\sim10^{-4}~\mathrm{Jy~km~s}^{-1}$ for the highest and lowest $\tau_{\mathrm{ff}}^{\mathrm{sc}}$, respectively.
    
    From the radially averaged flux from galaxies with different $\tau_{\mathrm{ff}}^{\mathrm{sc}}$ we see the similar trend as for varying $\varepsilon_\mathrm{SF}$ values. The flux profile from different model outcomes divides into distinguishable pairs for $\tau_{\mathrm{ff}}^{\mathrm{sc}} \leq 1$ and $\tau_{\mathrm{ff}}^{\mathrm{sc}} > 1$, although, the differences stop to be prominent at the galactic outskirts, where the flux is the weakest. Here, especially the profiles for $\tau_{\mathrm{ff}}^{\mathrm{sc}}=3$ and 5 get blended and cause major fluctuations by more than 2 orders of magnitude in the observed flux.
    
\subsection{Total galaxy emission}
    
    \begin{figure}[t]
    \resizebox{\hsize}{!}{\includegraphics{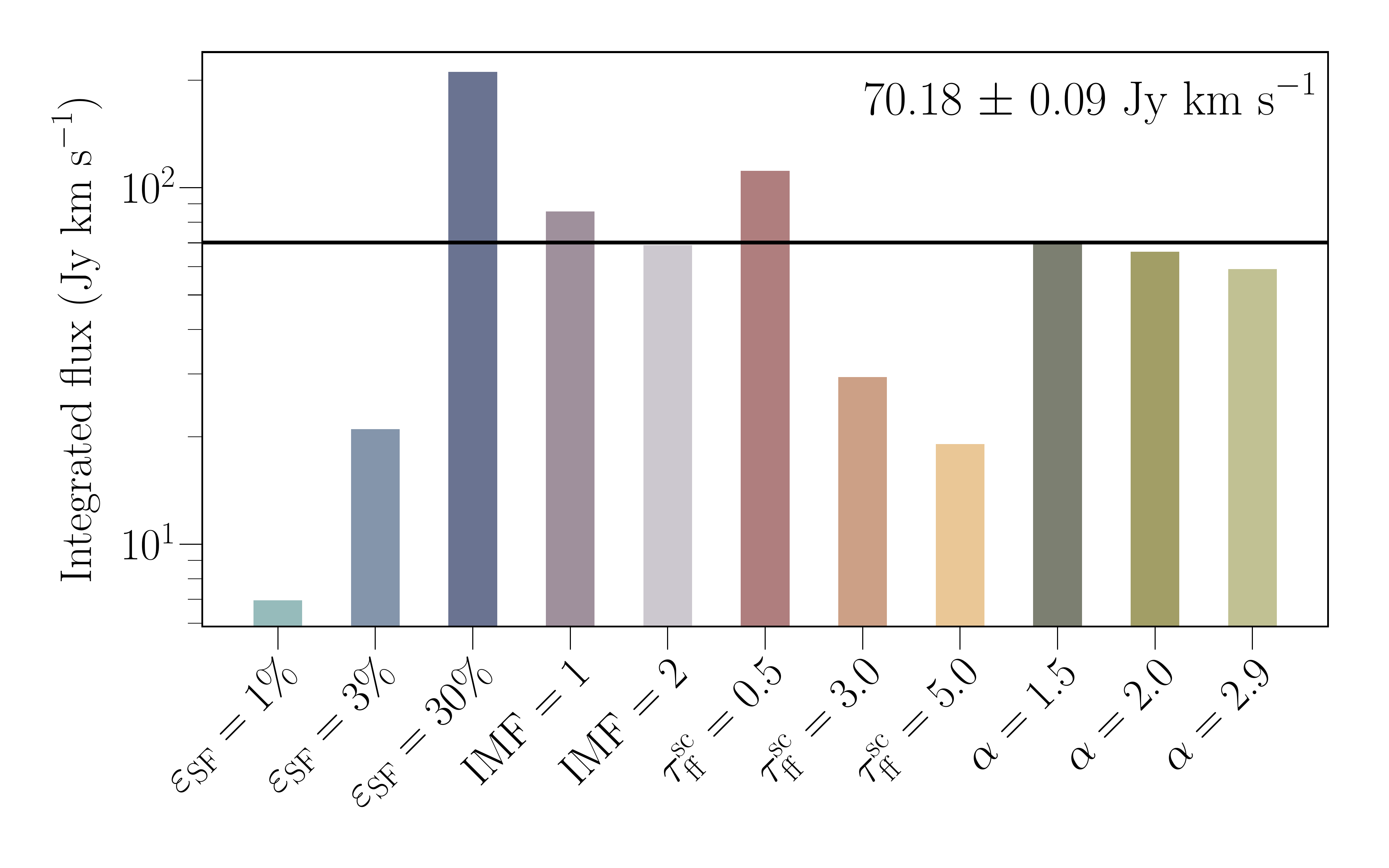}}
    \caption{The bar plot representing total galactic emissions derived from all of the clusters averaged over 10 simulations for each setup. The dashed black horizontal line corresponds to the standard setup described in Table \ref{tab:stand}.}
        \label{total}
    \end{figure}
    
    We calculated the integrated galactic emission for model outcomes with varying parameters (Fig. \ref{total}). The total integrated flux, $I_\mathrm{tot}$, was calculated from the mean flux distributions and for the standard setup is equal to $7.02\times10^{1}~\mathrm{Jy~km~s}^{-1}$. 
    
    From Fig. \ref{total} we see that only two $I_{\mathrm{tot}}$-values significantly exceed the default model outcome. The highest value of $I_{\mathrm{tot}}$ is observed for simulations with $\varepsilon_\mathrm{SF} = 30\%$ and is equal to $I_{\mathrm{tot}} = 2.11\times10^2~\mathrm{Jy~km~s}^{-1}$. The second highest value comes from the setup with  $\tau_\mathrm{ff}^\mathrm{sc} = 0.5$ with $1.12\times10^2~\mathrm{Jy~km~s}^{-1}$. For the varying $\alpha$ the highest total emission is derived for $\alpha = -1.5$ and falls almost at the same level as the output from the standard model. Similar thing happens for the top-heavy IMF, which exceeds the default $I_{\mathrm{tot}}$, by $1.56\times10^1~\mathrm{Jy~km~s}^{-1}$.
    
    The most visible changes are for the outputs that fall below the standard threshold. Here, we observe that the lowest total emission output is derived for the setup with the lowest $\varepsilon_\mathrm{SF}$ resulting in one order of magnitude drop in $I_\mathrm{tot}=6.96~\mathrm{Jy~km~s}^{-1}$. Subsequently, the second lowest value is a result of setting $\tau_\mathrm{ff}^\mathrm{sc}$ to 5.0 with $I_{\mathrm{tot}}=1.91\times10^1~\mathrm{Jy~km~s}^{-1}$. However, the second lowest value of $\varepsilon_\mathrm{SF}$ results in a very similar result with $I_{\mathrm{tot}}=2.10\times10^1~\mathrm{Jy~km~s}^{-1}$. Therefore, these two parameters have the biggest impact on emission and show the highest spread in derived $I_{\mathrm{tot}}$ values, while the lowest impact is observed for changes introduced to the molecular cloud mass distribution with the $\alpha$ index.

\section{Discussion}
\label{discussion}
    
    In the following, we will discuss model outcomes and their possible explanations. We will also evaluate the impact of different star-formation parameters and compare the joint effect of the most influential ones. Moreover, we will focus on addressing the question of what other star-formation-associated processes, not incorporated into the current version of the galaxy-in-a-box model, could influence the results. Finally, we will explore the implications of this proof-of-concept study for observations.
    
\subsection{Varying molecular cloud mass distributions}
    
    Molecular cloud mass distributions are usually described by a single power-law function (Eq. \ref{mcm}). Some studies \citep[e.g.,][]{1997ApJ...476..144M,2005PASP..117.1403R} propose truncated power-law distributions. However, when the truncated distribution applies, the cut-off point usually lies outside the mass range considered in this study, i.e., for $M_\mathrm{GMC} > 10^6 M_\odot$. The mass distribution can be expressed either in differential form, as in this work, or cumulative form with $\alpha > -1$ \citep{2015ARA&A..53..583H}. Many Galactic surveys report $\alpha > -2$  \citep[e.g.,][]{1987ApJ...319..730S,2010ApJ...723..492R,2014MNRAS.443.1555U}, while even steeper distributions are found in the outer parts of the Milky Way and in extragalactic regions, with $-2.9 < \alpha \leqslant -2$ \citep[e.g.,][]{2005PASP..117.1403R,2018MNRAS.477.5139G,2020ApJ...893..135M}. The $\alpha$ index indicates whether the majority of mass is contained in high-mass ($\alpha > -2$) or low-mass clouds ($\alpha < -2$). 

    We evaluated the impact of $\alpha$ on the predicted emission. It appears that steeper distributions result in lower medians and lower total fluxes (see Fig. \ref{alpha} \& \ref{total}). For the standard setup with $\alpha = -1.64$, we see a clear difference when comparing these outcomes to $\alpha = -2.9$. For these, the median values differ by $3.44\times10^{-4}~\mathrm{Jy~km~s}^{-1}$, with $\operatorname{IQRs}$ being narrower by $1.17\times10^{-3}~\mathrm{Jy~km~s}^{-1}$ for the latter one. This small, yet potentially observable level of discrepancy, means that the model could distinguish the molecular cloud distributions for slopes with a difference of the order of $\sim$1.
    
    This effect of lowered values with increasing steepness of the mass distribution is somewhat expected. Steeper distributions result in greater number of molecular clouds with smaller masses and produce smaller star-forming clusters. These greater number of low-mass clusters in turn emit less and thus lower total galactic emission, and this is what we see in Fig. \ref{alpha}.
    
    Comparing the impact of molecular cloud mass distribution and IMF, as these two seem to have the smallest impact on the predicted emission, we see that the standard and bottom-heavy IMFs result in median fluxes similar to molecular cloud mass distributions with $\alpha \geqslant -2$. However, the most bottom-heavy form of the molecular cloud mass distribution stands out, similarly to the top-heavy IMF. Therefore, when conducting observational comparisons to model outputs, putting constraints on the slope of $\alpha$, at least for its most varying values, or IMF shape, may be required to fine-tune the model and obtain better agreement with the observations. 
    
\subsection{IMF constraints}
    
    The parametrization of the IMF varies between studies, where the used format and high-mass cut-off differs between objects and redshifts \citep[e.g.,][]{2003PASP..115..763C,2008ApJ...675..163H,2008ApJ...674...29V,2020ARA&A..58..577S}, with the standard form being as follows: $N(M)\mathrm{d}M \propto M^{-2.35}$ \citep{1955ApJ...121..161S}. For more bottom-heavy IMF parametrizations, more low-mass stars are formed, while more top-heavy distributions lead to the presence of more high-mass stars.  
    
    In this study, we followed a widely used form of the IMF, the ``Chabrier IMF'' \citep{2003PASP..115..763C}, and adjusted it so it roughly represents the main three functional forms, i.e., standard, bottom-heavy, and top-heavy. As the building blocks of our model are molecular clouds from which individual star-forming clusters form, the IMF was directly influencing the stellar mass distribution of each cluster and emission components. By studying variations on these local, building blocks, and large galactic scales we see no significant variations imposed by the different IMF forms. However, for the standard IMF we see that the top-heavy distribution results in a slight increase in emission, while the opposite happens after adopting the bottom-heavy one. This result is expected. On the one hand, low-mass protostars dominate star formation in total mass and number \citep{2001MNRAS.322..231K}. The size of this population is increased or decreased for the bottom- and top-heavy IMFs, respectively. On the other hand, high-mass protostars are far more energetic than the low-mass ones. Moreover, with $\int I_{\nu}\propto {M_\mathrm{env}}^1$ water is a low-contrast mass tracer. Hence,  the more massive the envelope, the higher the emission.  
    
    When comparing results obtained for different IMF forms, we also see that the total flux obtained for the bottom-heavy IMF is very similar to the one derived for the standard one. These two are also very similar when we consider their flux distributions and radial profiles as seen in Fig. \ref{imf}. The same for their IQRs. Therefore, the model cannot distinguish these from one another. The top-heavy IMF, on the other hand, seems to differ when it comes to the IQR and range spanned by the flux distribution. However, the variation is in the range of $5.73-7.87\times10^{-3}~\mathrm{Jy~km~s}^{-1}$ for IQR and only $1.81-2.51\times10^{-3}~\mathrm{Jy~km~s}^{-1}$ for $\tilde{I}$. Nevertheless, this is the only IMF form that could be necessary to fine-tune the model when comparing it with observations.
    
    Looking at the total flux plot in Fig. \ref{total} we see that the output for the standard and bottom-heavy IMFs is comparable to other outputs derived for molecular cloud mass distributions for which $\alpha$ was set to $-1.5$ and $2.0$. The only difference between these setups can be seen in the shapes of their radial profiles, however, this may be not significant enough to distinguish these distributions from each other. 
    
    \begin{figure*}[t!]
    \resizebox{\hsize}{!}{\includegraphics[width=\textwidth]{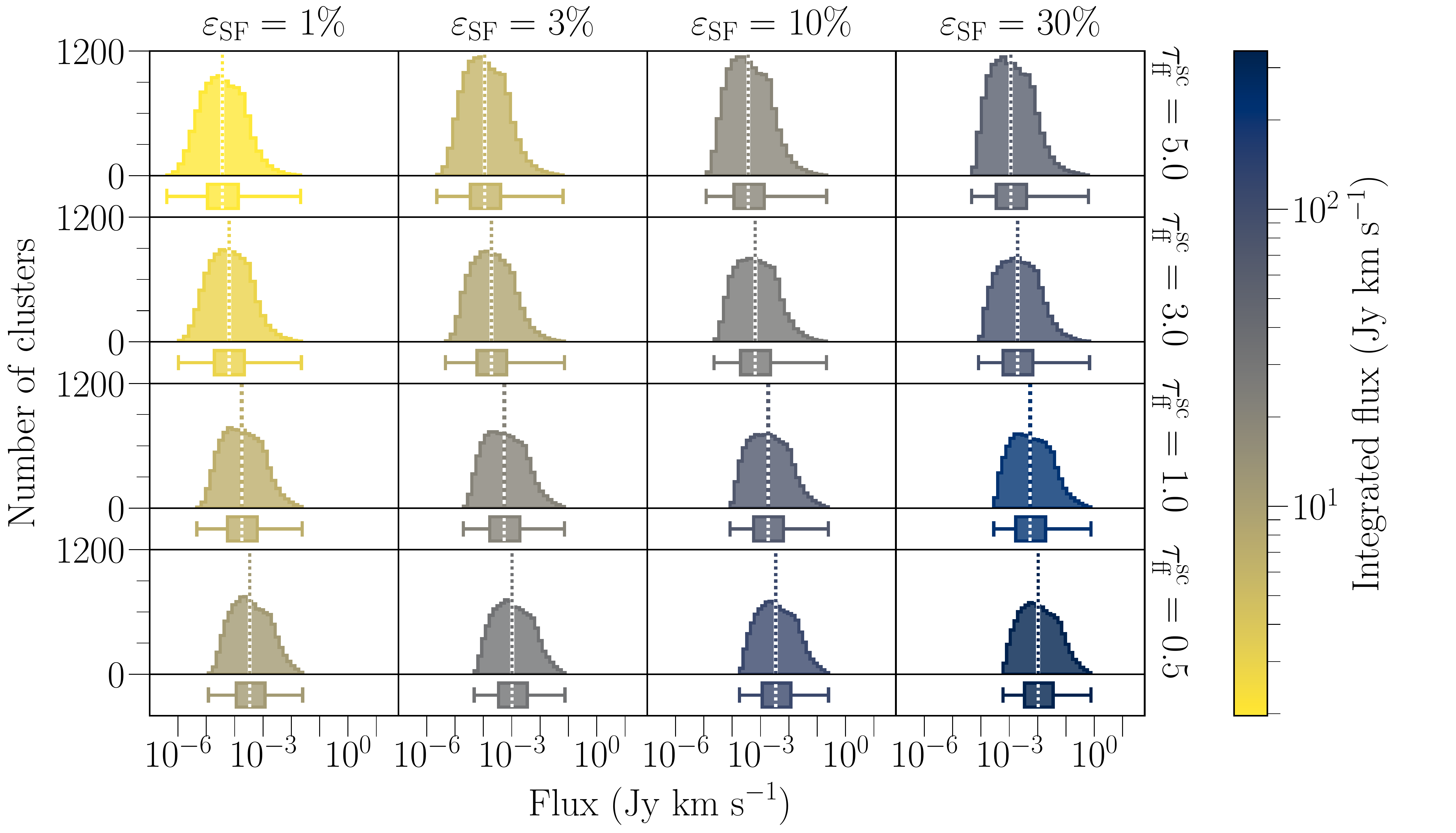}}
    \caption{Distributions of cluster emission derived for simulations where both the free-fall time scaling factor,$\tau_{\mathrm{ff}}^{\mathrm{sc}}$, and star formation efficiency, $\varepsilon_{\mathrm{SF}}$, varied. The vertical dashed lines correspond to the median flux of each distribution and a measure of central tendency in form of IQR is presented in the bottom box plots with whiskers spreading from the beginning  (0\%) to the end (100\%) of each distribution. These are the mean distributions from a series of 10 simulations for each varying pair of $\tau_{\mathrm{ff}}^{\mathrm{sc}}$ and $\varepsilon_{\mathrm{SF}}$. The y- and x-axes have the same ranges for all rows and columns. The color coding is based on the integrated fluxes of each distribution. The exact values of these fluxes are available in Fig. \ref{heatmap}.}
    \label{map_h}
    \end{figure*} 
    
\subsection{Effect of star formation efficiency}
    
    The star formation efficiency describes the amount of molecular gas that ends up in stars. The increase of $\varepsilon_{\mathrm{SF}}$ directly translates to an increase of the number of (proto)stars, which results in more emission from clusters. Different values of $\varepsilon_\mathrm{SF}$ are reported towards different types of sources across cosmic time, varying from $3\%$ in nearby molecular clouds to $30\%$ in Galactic embedded clusters \citep{2003ARA&A..41...57L} and extragalactic GMCs \citep{2019NatAs...3.1115D}. In this work, the impact of $\varepsilon_\mathrm{SF}>30\%$ is not evaluated, as $\varepsilon_\mathrm{SF}$ is closely related to the gas depletion time and with higher $\varepsilon_\mathrm{SF}$, molecular gas is used at a higher rate and is sustained for a shorter time.  
    
    Analyzing the impact of $\varepsilon_{\mathrm{SF}}$ on the expected emission locally and on a galactic scale, we observe a clear and systematic increase of emission with increasing $\varepsilon_{\mathrm{SF}}$. The observed increase in emission is roughly proportional to the increase in $\varepsilon_{\mathrm{SF}}$. There is a shift of the flux distributions as seen in Fig. \ref{sfe}. The $\operatorname{IQRs}$ follow the same trend and vary between $\sim6\times10^{-4}-2.0\times10^{-2}~\mathrm{Jy~km~s}^{-1}$. This suggests that the model can be used to distinguish different values of $\varepsilon_{\mathrm{SF}}$, at least when no other parameter varies. 
    
    Distributions drawn from model outputs with varying $\varepsilon_{\mathrm{SF}}$ show significant variations when considering all of the analysis, which is also true for the impact of $\tau_{\mathrm{ff}}^{\mathrm{sc}}$. However, these two parameters significantly differ when it comes to the shape of the flux distributions and radial profiles. Therefore, it should be possible to evaluate which parameter could explain the observed galactic emission.
    
    \begin{figure*}[t!]
    \resizebox{\hsize}{!}{\includegraphics[width=\textwidth]{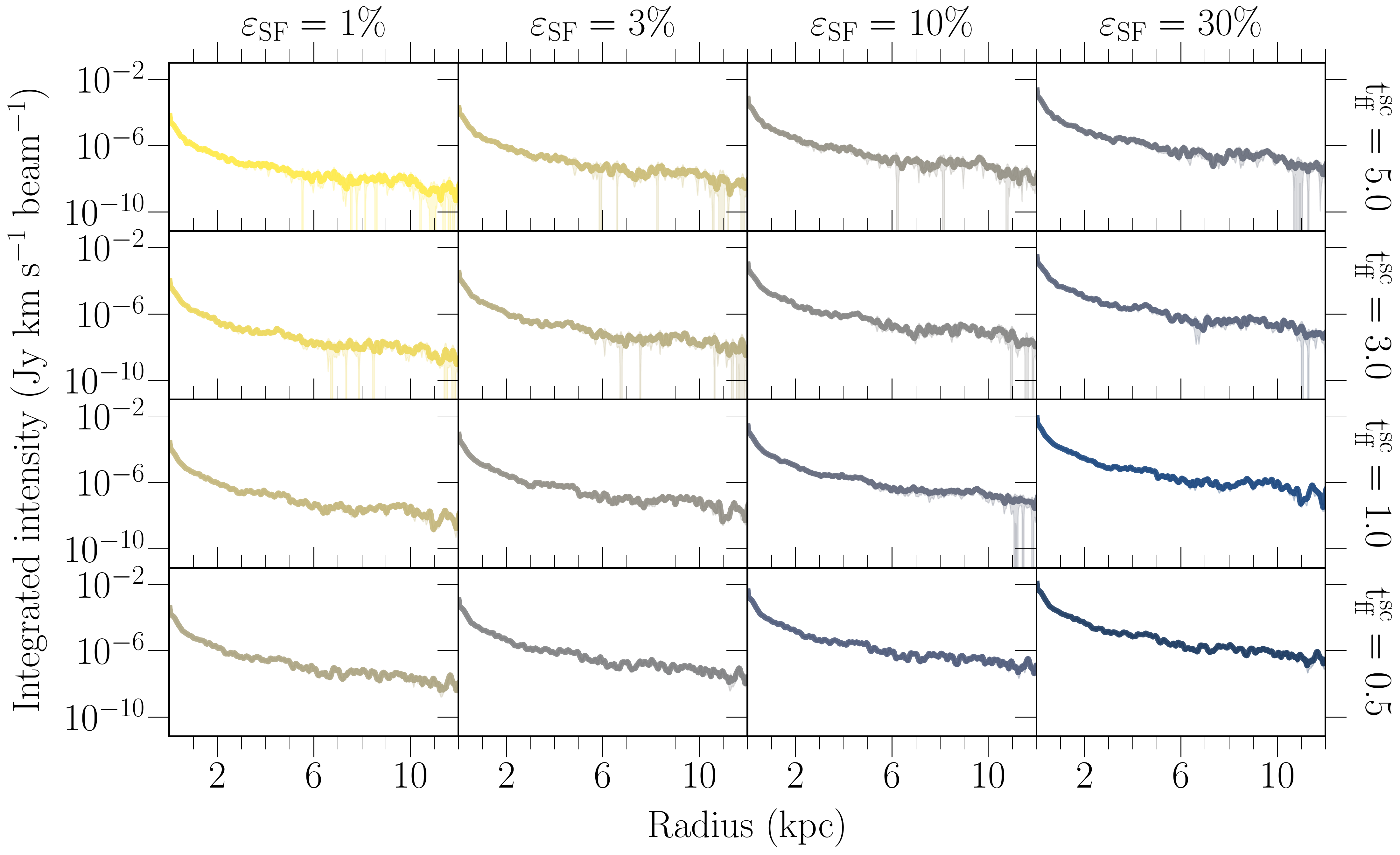}}
    \caption{Radial profiles of emission from the same set of simulations as in Fig. \ref{map_h}. Solid lines correspond to the mean radial profile, while the shaded regions represent the variability of each profile, based on the standard deviation of each profile. The color coding follows the one from Fig. \ref{map_h}.} \label{map_r}
    \end{figure*}
    
\subsection{Influence of the free-fall time scaling factor}
    
    The last considered parameter is the free-fall time scaling factor, $\tau_{\mathrm{ff}}^{\mathrm{sc}}$. Here, we arbitrarily chose all of the values to explore how altering the ages of clusters could affect the expected emission. With $\tau_{\mathrm{ff}}^{\mathrm{sc}} < 1$, we effectively lower the ages of protostars within the cluster and therefore increase the contribution from Class 0/I. Therefore, with lower $\tau_{\mathrm{ff}}^{\mathrm{sc}}$ values we would expect higher emission both globally and locally.
    
    From the flux distributions and radial profiles in Fig. \ref{eff} we see that there is indeed an increase of flux with the decrease of $\tau_{\mathrm{ff}}^{\mathrm{sc}}$. Moreover, all of the distributions tend to flatten with this decrease. We also observe a peculiar shape of the distribution derived for the smallest $\tau_{\mathrm{ff}}^{\mathrm{sc}}$. The possible explanation for this peculiar shape is that such a huge change in free-fall time results in constraints on the age distribution of clusters within galaxies. It is also the distribution with the higher median, which indicates a greater number of Class 0 sources within clusters, which produce more molecular emission from outflows.
    
    Following \cite{2018A&A...618A.158K}, the fraction of Class 0/I cores decreases with the age of the cloud and reach a steady-state at $\sim0.5$ Myr. Therefore, as the scaling of the free-fall time increases, especially when  $\tau_\mathrm{ff}^\mathrm{sc} \geqslant 1$, clusters more accurately represent the dynamics of stellar formation. This in turn results in a greater range of flux distributions and lower median fluxes, as the fraction of Class 0/I cores decreases. 
    
    The outcome for $\tau_{\mathrm{ff}}^{\mathrm{sc}}=5.0$ is similar to the one for $\varepsilon_\mathrm{SF}=3\%$, when considering the cumulative galactic flux as seen in Fig. \ref{total}. Nevertheless, the difference between these outputs is potentially observable, especially that $\tau_{\mathrm{ff}}^{\mathrm{sc}}=5.0$ gives a flatter flux distribution. Therefore, the model could distinguish the emission for these global parameters.
    
\subsection{Interplay of the most influential parameters}
    
    The most influential parameters in the model are $\tau_{\mathrm{ff}}^{\mathrm{sc}}$ and $\varepsilon_\mathrm{SF}$. Thus, to understand and explore the combined effect of these parameters on simulated galaxies we run the model for all of the possible combinations of the values considered in this study. Then, we evaluated the outcomes of these simulations by calculating the distributions of cluster fluxes and their corresponding midspreads (see Fig. \ref{map_h}) and galactic radial profiles (Fig. \ref{map_r}). Moreover, we color-coded the results of each simulation based on the integrated intensities of the flux distribution. The heat map with corresponding integrated fluxes is presented in Appendix \ref{appendix:a}.
    
    The distribution of fluxes changes accordingly to what we observed when studying the impact of $\tau_{\mathrm{ff}}^{\mathrm{sc}}$ and $\varepsilon_\mathrm{SF}$ separately, namely that median flux and integrated intensity within galaxies increases with increasing $\varepsilon_\mathrm{SF}$ and decreasing $\tau_{\mathrm{ff}}^{\mathrm{sc}}$. Interestingly, $\varepsilon_\mathrm{SF}$ seems to mainly influence the median flux by shifting the distribution towards higher flux values proportionally to the magnitude of the increase. Also, the shift is not associated with any significant changes in the shape of the distributions. On the other hand, $\tau_{\mathrm{ff}}^{\mathrm{sc}}$ increases median fluxes but does not shift the whole distribution. What happens is that with the decrease of $\tau_{\mathrm{ff}}^{\mathrm{sc}}$ the distributions flatten and, based on the midspreads, the high-flux tail seems to catch up with the rest of the distribution. Subsequently, there is a decrease in the spread of observed flux values. The lower-flux part of the distribution ``shifts'' towards higher flux values, but it does not affect the highest flux values.
    
    The changes observed on galactic scales also reveal complex outcomes of the interplay of these parameters. Here we observe that $\varepsilon_\mathrm{SF}$ basically scales the radial profiles up and increases the level of each emission point, especially in the inner galaxy where most of the clusters reside. It also influences the visibility of the spiral arm ``bumps'' in the radial profiles. Surprisingly, these bumps are more prominent with the increase of the free-fall time scaling factor. However, this change is also associated with the increased radial profile variability.
    
    By looking at the simulations obtained for all of the available combinations, we see that the impact of each parameter is different, and the only common characteristic is a strong influence on the observed emission. From flux distributions, we can see that with spatially resolved observations, one could estimate the possible value of each parameter because they introduce very distinct features to the shape and properties of each distribution. While in case of unresolved observations, one could try to evaluate these values based on the features seen in the galactic radial profiles. Therefore, our model can be used to unveil these global star formation parameters or at least indicate which one has the prevalence in increased or decreased star formation activity in a galaxy.

\subsection{Other effects}

    More things could play a role in water excitation. These include local and global effects of star-formation processes and galactic evolution and structure. 
    
    The warm ($\gtrsim 100 \mathrm{K}$) and dense ($\gtrsim 10^6 \mathrm{cm}^{-3}$) inner regions of protostars, the so-called hot cores, exhibit conditions that support the presence of a rich chemistry \citep{2009ARA&A..47..427H}. Under such conditions, all water ice should sublimate, and the observed gaseous water abundances should match expected water ice abundances. However, observations do not follow these expectations showing that most of the observed gaseous water is universally associated with warm outflowing and shocked gas, with a negligible contribution from hot cores  \citep{2021A&A...648A..24V}. Indeed, the low surface brightness of the hot cores along with the high dust opacity at 1 THz obscuring the hot cores makes them practically invisible in a \textit{Herschel} beam \citep{2013ApJ...769...19V,2012A&A...542A..76H}. 
    
    On larger scales, the question arises about the emission from molecular clouds themselves. Here, water vapor is generally not detected in the Milky Way \citep[e.g.,][]{2020A&A...641A..36D}. The only noteworthy exception would be regions exposed to enhanced UV radiation, the so-called Photon Dominated Regions with one narrow emission component \citep{2012A&A...546A..29B}. However, overall molecular cloud contribution to the observed water emission is insignificant for the results of this study, particularly for the higher-excited 2$_{02}$--1$_{11}$ transition.
    
    Galaxy-wise, Active Galactic Nuclei (AGNs) could play a role in water emission enhancement or decrease, both locally and globally. Studies report quenching of star formation in AGN host galaxies, which would lower the number of protostars and thus outflows \citep[e.g.,][and references therein]{2012ARA&A..50..455F,2015ARA&A..53..115K,2021A&A...648A..24V}. Moreover, AGNs can produce conditions favoring molecular excitation or dissociation if the radiation gets too strong. The exact influence of the AGN feedback on water excitation is not well understood, but it appears that AGN presence has little impact on excitation of the water line considered in this study, i.e., para-H$_2$O \(2_{02} - 1_{11}\) line at 987.927 GHz. Specifically, \citet{2019ApJ...880...92J} spatially resolved H$_2$O emission in this transition toward the Cloverleaf quasar, which is a strongly lensed AGN, at a resolution of 1 kpc, but found no trend with distance to the actual AGN. Thus, considering AGN feedback would likely have a negligible effect on the results of this study.

\subsection{Implications for observations}
    
    Verification of the model can only be obtained by benchmarking its outcomes against observations. Ideally, these observations should spatially resolve individual star-forming clusters. This way, the cluster flux distribution is compared with a simulated galaxy. To come down to $\sim10$ pc scales and spatially resolve the whole range of molecular cloud sizes, the resolution should be of the order of 0\farcs3 at 7.6 Mpc. 

    The results presented from our proof-of-concept study are for a resolution of 2\farcs55, which at 7.6 Mpc corresponds to $\sim70$ pc. This resolution is comparable to the resolution at which M51 was observed as part of the PAWS program \citep{2013ApJ...779...42S}, and where individual GMCs are resolved. Therefore, smaller clouds are unresolved in the galactic image. However, only a handful of high-redshift star-forming galaxies are spatially resolved in H$_2$O emission, although then at a resolution of $\sim$ 1 kpc-scales \citep{2019ApJ...880...92J}. Most observations do not resolve emission, and comparisons would have to be done based on the total fluxes or water line luminosities, rather than on radial profiles or shape of cluster flux distributions. With this assumption, we can make a tentative comparison of water line luminosities observed towards nearby and distant galaxies with the ones derived in this study. 
    
    The average total flux of $\sim70~\mathrm{Jy~km~s}^{-1}$, corresponding to $\sim1300~\mathrm{L}_\odot$, derived for the simulated galaxies in this study remains $\sim$ one order of magnitude below the luminosity derived for the nearby starburst M82 \citep{2013yang}, which is not surprising considering that M82 has $\sim$ one order of magnitude higher SFR \citep[e.g.,][]{2001grijs} than the Milky Way or M51. The observed luminosities towards several LIRG/ULIRGs at larger distances \citep{2013yang} or high-$z$ starbursts at $z\sim2-4$ \citep[e.g.,][]{2011ApJ...741L..38V,2011A&A...530L...3O,2013A&A...551A.115O,2016A&A...595A..80Y,2019ApJ...880...92J} remain up to $\sim 2-4$ orders of magnitude higher. However, this difference is expected and consistent with the increasing SFRs of these galaxies, especially when considering the high-$z$ ones where SFRs often exceed $\sim1000~\mathrm{M}_\odot/\mathrm{yr}$, which naturally boosts star formation, and hence the emission coming from the protostellar outflows. However, more comparisons are needed to fully assess the differences between the model and high-redshift observations, but this is beyond the scope of this paper.
    
    There are several ways in which to interpret the difference between the model outcomes and the observations of high-$z$ galaxies.
    First of all, our template galaxy resembles the nearby M51 galaxy. We chose this particular galaxy instead of a well-studied high-redshift galaxy because we wanted to start with an object with a known molecular-cloud distribution \citep[e.g.,][]{2013ApJ...779...46H,2014ApJ...784....3C}, as this is one of the building blocks in our model. Second, our results are for a standard IMF \citep{2003ARA&A..41...57L}; there are indications that IMFs toward high-$z$ galaxies are significantly more top-heavy even than what we tested here, which would serve to further boost emission from the high-mass stars. However, this, in turn, implies that we are dealing with a different spatial setup, galactic size, and internal galactic environment. This size difference is very prominent, as spatially-resolved high-redshift galaxies have radii in the range of $0.95-2.24$ kpc \citep{2019ApJ...880...92J}, while M51 has a radius of $\sim12$ kpc. 
    
    On the other hand, there is a reasonable agreement between the model results and observations of galaxies that lie closer to M51. \cite{2021A&A...647A..86S} reported water flux measurements from the \textit{Herschel} SPIRE observations towards the NGC 1365 galaxy, lying at a distance of 18.6 Mpc \citep{1998Natur.395...47M}. The observed flux corresponds to $3081.9~\mathrm{Jy~km~s}^{-1}$, which falls on the higher end of the fluxes derived for the model results when distance-corrected, and if $\sim$ one order of magnitude difference in the SFR between the Milky Way/M51 and NGC 1365 would be taken into account. For a nearby starburst, Mrk 231 at a distance of $\sim200$ Mpc \citep{2010A&A...518L..42V}, \cite{2011A&A...530L...3O} reports a flux of $718~\mathrm{Jy~km~s}^{-1}$, which distance- and SFR-corrected also falls on the high end of the simulated fluxes. 
    
    It is clear that both the star-formation efficiency and the free-fall scaling parameter can affect the H$_2$O flux dramatically (e.g., Fig. \ref{map_h}). A single integrated H$_2$O flux is not going to constrain either parameter, and additional constraints are needed. To constrain the star-formation efficiency, for example, the total number of stars formed combined with the amount of molecular material available should be observed. The former is best constrained through an in-depth look into stellar masses in galaxies, both nearby and at high-redshift. One way to do it is through near- and mid-IR observations, where the James Webb Space Telescope (JWST) will provide a great advance, especially for the high-redshift regime. The molecular material available can be probed either through low-$J$ CO emission or dust emission. Although there are known problems with both tracers \citep[e.g.,][]{pitts2021}, they are the best tracers at the moment for the total gas mass. Thus, with the combination of JWST observations of the stellar mass and, e.g., ALMA observations of the total gas mass, the star formation efficiency can be independently constrained.

    Another thing to consider could be the detailed comparisons of spatially resolved observations with model results, where it would be possible to evaluate which sets of the star-formation parameters can reproduce the galactic emission. Here, for example, by analyzing the flux distribution of the observed emission (similar to Fig. \ref{map_h}), it would be possible to put constraints on these parameters and pinpoint their most probable values.

\section{Conclusions}
\label{conclusions}
        
    We created a galactic model of emission that could arise from active galactic star-forming regions. In this paper, we demonstrated the main principles behind the galaxy-in-a-box model and explored how it can serve as a tool to study and better understand star-formation activity in galaxies even at high redshift. For a template galaxy set to resemble the
    grand-design spiral “Whirlpool Galaxy” M51, we evaluated the impact of important global star-formation parameters on model results. We conducted this parameter space study for the para-H$_2$O \(2_{02} - 1_{11}\) line at 987.927 GHz. The main results are as follows:
    \begin{itemize}
    \item  emission from the para-H$_2$O \(2_{02} - 1_{11}\) line is a low-contrast tracer of active star formation with $\int I_{\nu}\propto {M_\mathrm{env}}$;
    \item the initial mass function along with molecular cloud mass distribution have little impact on predicted water emission;
    \item increase/decrease of star formation efficiency, $\varepsilon_{\mathrm{SF}}$, increase/decrease the predicted emission, both locally and globally; 
    \item with the decrease of free-fall time scaling factor, $\tau_{\mathrm{ff}}^{\mathrm{sc}}$, we observe a corresponding increase in galactic emission and flattening of star-forming flux distribution, which indicates increasing populations of Class 0 and Class I protostars; 
    \item at the moment, further constraints are needed to break model degeneracies; these additional constraints include JWST observations combined with low-$J$ CO observations, and resolved observations of H$_2$O emission. 
    \end{itemize}

    A tentative comparison of model outcomes with observational data for high-redshift galaxies yields realistic results and opens new paths to improve the model, so it can become a reliable proxy to unveil star formation in galaxies throughout cosmological times. In the near future, we plan to: (i) introduce the possibility to turn on/off AGN feedback and (ii) conduct detailed comparisons of model results with observations of local and distant LIRGs, ULIRGs, and HyLiRGs. Furthermore, since our model is not designed specifically for water molecules, we intend to explore the results for other unique outflow tracers, like high-$J$ CO ($J \ge 10$). It will be important to constrain which global star-formation parameters that have not impacted our results for water emission will behave differently for other molecular tracers. 

\begin{acknowledgements}
The research of KMD and LEK is supported by a research grant (19127) from VILLUM FONDEN. This article has been supported by the Polish National Agency for Academic Exchange under Grant No. PPI/APM/2018/1/00036/U/001.
\end{acknowledgements}

\bibliographystyle{aa}
\bibliography{references}
\appendix
\section{Integrated flux values of flux distributions}
\label{appendix:a}
\begin{figure}[t!]
\resizebox{\hsize}{!}{\includegraphics{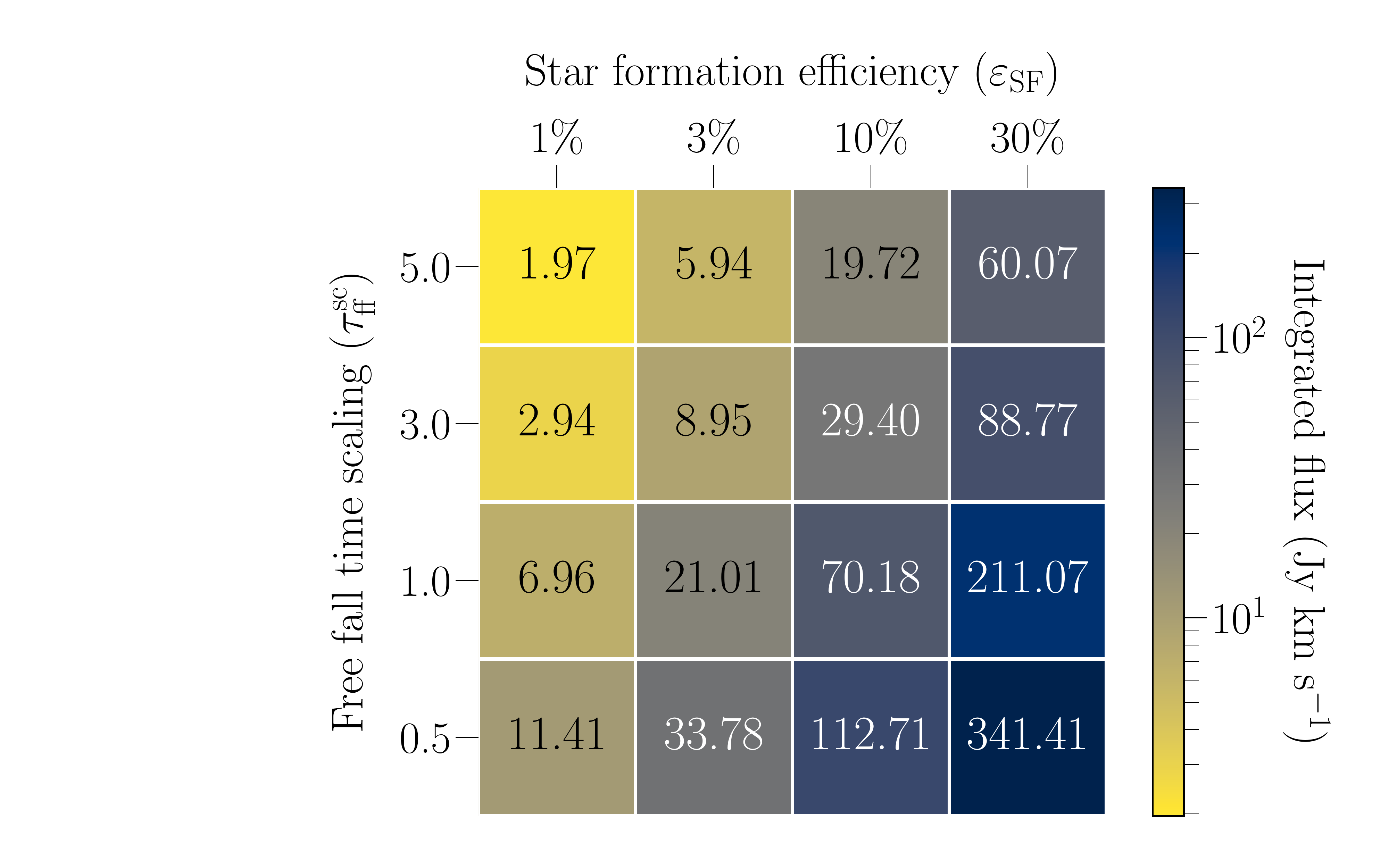}}
\caption{The heat map with integrated flux values calculated from the flux distributions presented in the Fig. \ref{map_h}.}
\label{heatmap}
\end{figure}

In order to evaluate the interplay of the most influential parameters, i.e., $\tau_{\mathrm{ff}}^{\mathrm{sc}}$ and $\varepsilon_{\mathrm{SF}}$, we color coded the results from each set of simulations based on the integrated flux values calculated from the cluster flux distributions. Also, we created a corresponding flux value guide in a form of a heat map presented in the Fig. \ref{heatmap}.

\end{document}